\PassOptionsToPackage{svgnames}{xcolor}
\documentclass[11pt,reqno]{article}

\usepackage[numbers,sort&compress]{natbib}
\usepackage{amsthm, amsmath, amsfonts, amssymb, amscd, mathtools, youngtab, euscript, mathrsfs, verbatim, enumerate, multicol, multirow, bbding, color, babel, esint, geometry, tikz, tikz-cd, tikz-3dplot, tkz-graph, array, enumitem, thm-restate, thmtools, datetime, graphicx, tensor, braket, slashed, standalone, pgfplots, ytableau, subfigure, wrapfig, dsfont, setspace, wasysym, pifont, float, rotating, adjustbox, pict2e,array, physics, pgfplots}
\usepackage[colorlinks=true,linktocpage=true,linkcolor=DarkBlue,citecolor=DarkBlue,urlcolor=DarkBlue]{hyperref}
\usepackage{amsfonts}
\usepackage{longtable}
\usepackage{tabularx}
\usepackage{caption}
\usepackage{booktabs}
\usepackage{amsmath}
\usepackage{etoolbox}
\usepackage{makecell}
\usepackage{blindtext}
\usepackage[T1]{fontenc}
\usepackage[utf8]{inputenc}
\usepackage[]{authblk}

\newcommand{\beq}{\begin{equation}}
\newcommand{\eeq}{\end{equation}}
\newcommand{\Mpl}{{M_{\rm pl}}}
\newcommand{\Vc}{{\mathcal{V}}}

\title{TCC in the interior of moduli space\\
and its implications for the string landscape and cosmology}
\author[a,b]{Alek Bedroya}
\author[c,d]{Qianshu Lu}
\author[b,e]{Paul Steinhardt}
\affil [a]{\small Princeton Gravity Initiative, Princeton University, Princeton, NJ 08544, USA}
\affil[b]{Jefferson Physical Laboratory, Harvard University, Cambridge, MA 02138, USA}
\affil[c]{School of Natural Sciences, Institute for Advanced Study, Princeton, NJ 08540, USA}
\affil[d]{Center for Cosmology and Particle Physics, Department of Physics, \protect\\ New York University, New York, NY 10003, USA}
\affil[e]{Department of Physics, Princeton University, Princeton, New Jersey 08544, USA}

\date{}

\begin{document}

\maketitle
\normalsize
\begin{abstract}

We consider the classical Friedmann-Robertson-Walker solutions that describe a
universe undergoing a transition from an accelerating expansion phase in the past
to an eternal decelerating expansion phase in the future, driven by a scalar field
evolving in a potential energy landscape. We show that any solution for which the accelerating phase violates the Trans-Planckian Censorship Conjecture (TCC), even in the interior of moduli space, never approaches the asymptotic vacuum with zero particles. Based on the assumption that the effective field theory must be valid for the vacuum on the asymptotic boundary, as motivated by holography and string theory, we argue that (multi-field) scalar potentials with such solutions are disallowed, thus strengthening the case for TCC.  In particular, the results imply a new set of complex and highly-nonlinear constraints across the entire string landscape which may make realizing inflation impossible.

  
\end{abstract}
\newpage
\tableofcontents

\section{Introduction}

One of the special features of quantum gravity is the emergence of spacetime from boundary observables. Such observables have the benefit of not relying on the spacetime topology, which can fluctuate in quantum gravity. Scattering amplitudes in Minkowski space and boundary correlation functions in AdS space \cite{Gubser:1998bc,Maldacena:1997re,Witten:1998qj} are examples of such boundary observables. Note that here by boundary we refer to the boundary of the compactified spacetime after a conformal transformation. It is desirable to extend this logic to Friedmann-Robertson-Walker (FRW) backgrounds given their relevance for cosmology. However, the task of defining such boundary observables in FRW backgrounds is not straightforward. Boundary observables are conventionally defined by taking the limit of  appropriate bulk correlation functions in a chosen vacuum. In this work, we consider cases where there is a natural choice of vacuum and examine their implications. 

For example, in perturbative descriptions of string theory,  there can be small perturbations about the vacuum in the infinite past or future. This statement also holds for time-dependent backgrounds that satisfy loop-corrected equations of motion \cite{Fischler:1986ci,Fischler:1986tb,Callan:1988wz}. Given that the scalar potential in string theory is always generated by non-zero string coupling \cite{Fischler:1986ci}, FRW backgrounds in string theory are time-dependent solutions to the $g_s$-corrected equations.  The typical {\it expanding} FRW solutions realized in string theory are driven by moduli scalar fields evolving down a potential energy landscape where these solutions  become singular in the infinite past and have power-law decelerated expansion in the future \cite{Bedroya:2022tbh}. In this case, the natural choice of boundary vacuum to consider for our analysis is the future vacuum. 

Beginning from this vacuum  defined on the asymptotic boundary of spacetime, we identify a class of theories (defined by their scalar field potentials) that are inconsistent with effective field theory (EFT), and hence, with  quantum gravity. For this class of theories, we find that EFT  breaks down if the Trans-Planckian Censorship Conjecture (TCC) \cite{Bedroya:2019snp} is violated. Motivated by studies of string theory, TCC postulates that quantum gravity prohibits long periods of accelerated expansion that can stretch fluctuations with  Planckian wavelengths to Hubble sized wavelengths. This conjecture is well supported  in string theory in the asymptotic regime of  moduli space where there is power-law decelerated expansion at future infinity (see \cite{Shiu:2023nph,Shiu:2023fhb} for a discussion of asymptotic behavior of scalar field cosmologies for multi-field potentials). For example, in this asymptotic regime, the validity of TCC has been justified in earlier work by one of us \cite{Bedroya:2022tbh} based on the existence of boundary observables at future infinity. However, this earlier analysis has limited phenomenological implications because it  only constrains potentials in the large field/far future limit. For example, a very long-lived quasi-de Sitter space in the past that smoothly evolves into an eternally decelerating universe is not ruled out by this analysis. Although the bulk of the evidence for TCC comes from the asymptotic behavior in the string landscape, there are also independent arguments that non-trivially relate TCC to other Swampland conjectures. For example, the authors in \cite{vandeHeisteeg:2023uxj} showed that the emergent string conjecture \cite{Lee:2019wij}, which is independently motivated \cite{Bedroya:2024ubj,Basile:2023blg}, has similar implications as TCC for the slope of monotonic scalar potentials in the interior of the moduli space. In this paper we strengthen the case for TCC by providing a bottom-up argument that precisely reproduces TCC for a wide class of potentials with less assumptions.

This paper is organized as follows. In Sec. \ref{Sec:summary}, we summarize the arguments that underlie our conclusions.  In Sec. \ref{Sec:2}, we show that, if TCC is violated by a prolonged period of accelerated expansion, the initial state must have had trans-Planckian excitations inconsistent with EFT to evolve into the future vacuum. We first demonstrate this for a toy model where the quasi-stationary modes and particle production can be calculated analytically. Then we provide a general argument in terms of the WKB approximation. Perhaps, the most surprising finding is that we show that no matter how smoothly the expansion transitions from TCC-violating accelerated expansion to an eternally decelerating expansion, the past and future vacua will not evolve into one another. We explain why our conclusion does not violate the adiabatic theorem. In Sec. \ref{Sec:curvature}, we extend our arguments to a class of spatially-curved open universes to show that our findings are robust against sufficiently small classical perturbations. 

In Sec. \ref{sec:black hole}, we discuss the difference between our argument presented in this paper and the "trans-Planckian problem" \cite{Martin:2000xs,Brandenberger:2012aj,Jacobson:1991gr} raised decades ago in the context of inflationary cosmology and black hole physics. We will show that our argument does not imply any inconsistency when applied to trans-Planckian blueshifts near black hole horizons, although it imposes significant constraints on accelerated expansion in cosmology.

We discuss the implications of our results for the critical points of the scalar potential in Sec. \ref{Sec:3}. Since our argument relies on perturbations around a given classical background, it falls short of bounding metastable quasi-de Sitter vacua which decay non-perturbatively. However, we can rule out long-lived unstable quasi-de Sitter phases that violate TCC. 

In Sec. \ref{Sec:5}, we consider the implications for inflationary cosmology. 
 We argue that, in a typical string landscape described by a multiplicity of  scalar fields, the extended TCC presented here imposes extraordinarily complex and highly-nonlinear constraints across the entire string landscape that may not be possible to satisfy.


\section{Summary of the argument}\label{Sec:summary}

In this section, we explain our assumptions and summarize our argument. Our goal is to determine which scalar field theories are consistent with quantum gravity generally. To achieve this, we study the set of classical FRW solutions that describe a universe that undergoes a transition from an accelerating expansion phase in the past to a decelerating expansion in the future, driven by the scalar field evolving along its potential. In order to rule out a theory, we make an assumption that is motivated by holography and is satisfied in perturbative string theory: namely, we assume that the EFT is valid in the state with zero particle excitations in the future. In perturbative string theory, which is a theory of scattering amplitudes, one can define a Hilbert space in terms of particle excitations around an asymptotic vacuum. For such a definition to make sense, we need at least one asymptotic vacuum. The EFT is then defined to approximate the low-energy physics for such states in the Hilbert space. Therefore, this condition is satisfied by definition. Moreover, since expanding FRW solutions realized in string theory are typically singular in the past, the only available boundary is future infinity and the future vacuum is the only natural ground state from the perspective of holography. We will show that a certain subset of TCC-violating solutions cause the EFT to breakdown when describing this ground state, thus these solutions cannot be consistent with string theory. We cannot rule out all TCC-violating solutions with our present arguments; the conditions for our argument and some notable cases where our argument does not apply are listed below. However, our bottom-up argument has the advantage of exactly reproducing TCC both qualitatively and quantitatively in the interior of the moduli space for a wide range of potential energy landscapes.

In the above discussion, we have explained how certain combinations of a classical solution and a natural choice of vacuum state can be shown to be inconsistent with quantum gravity. If even one inconsistent solution can be found for a (multi-field) scalar potential, the potential is ruled out even if for the same potential, there also exist combinations of classical solutions and vacua that are consistent. The idea that every time-evolution of a valid theory of quantum gravity has to satisfy certain consistency conditions is a familiar idea in the Swampland program, which aims to identify those conditions for the quantum gravity landscape. TCC and Penrose inequality are two such examples which are conjectured consistency conditions for all solutions in a theory of quantum gravity \cite{Bedroya:2019snp,Folkestad:2022dse}. However, we will not rely on any of the Swampland conjectures in our argument and give a bottom-up argument for TCC.

The TCC-violating solutions of interest are ones  with the following properties:
\begin{enumerate}
\item \textbf{TCC-violating accelerated expansion phase: } an epoch or  sequence of epochs of accelerating expansion that collectively last long enough to stretch modes with Planckian wavelength to super-Hubble wavelengths. 
\item \textbf{Transition at finite time to a long decelerated expansion phase}: the decelerated expansion phase must last long enough that the modes whose wavelengths went from sub-Hubble to super-Hubble during the acceleration phase re-enter the Hubble horizon and become sub-Hubble during the decelerated expansion phase, and remain sub-horizon in the asymptotic future. 
\item \textbf{Small non-positive initial spatial curvature:} If $\Omega_k$ is precisely zero, the decelerated ($\ddot{a}<0$) expansion phase must be eternal. For non-zero $\Omega_k$, the decelerated expansion phase can transition to an expanding phase in which $\Omega_k$ does not vanish at future infinity. In such backgrounds, the acceleration vanishes at future infinity ($\ddot a\rightarrow 0$). Here $\Omega_k$ is the ratio of the spatial curvature to the critical density. In the latter case, the initial spatial curvature must have a magnitude small enough that the non-accelerating phase begins after Planckian wavelength modes stretched to super-Hubble wavelengths created during the accelerating phase have re-entered the horizon during the decelerating expansion phase.
\item \textbf{Semi-classical evolution:} the entire cosmological background evolution from the initial accelerating phase to the future vacuum in the asymptotic limit of the eternally expanding phase(s) that follow  is well-described  by the classical equations of motion.   For example, this eliminates cases in which  quantum-generated bubble nucleation or large quantum fluctuation effects are essential/dominant in transitioning from the accelerating to the decelerating expansion phase.  
\end{enumerate}

If a classical solution with these properties exists, then our arguments show that the EFT must break down in the past for the state that evolves into the future vacuum.

\begin{figure}[H]
    \centering

\tikzset{every picture/.style={line width=0.75pt}} 

\begin{tikzpicture}[x=0.75pt,y=0.75pt,yscale=-1,xscale=1]

\draw   (198,20) -- (504,300) -- (198,300) -- cycle ;
\draw    (198,185) .. controls (247,189) and (399,226) .. (504,300) ;
\draw  [color={rgb, 255:red, 208; green, 2; blue, 27 }  ,draw opacity=1 ] (183,186) .. controls (178.33,186.04) and (176.02,188.39) .. (176.06,193.06) -- (176.41,231.56) .. controls (176.47,238.23) and (174.17,241.58) .. (169.5,241.63) .. controls (174.17,241.58) and (176.53,244.89) .. (176.59,251.56)(176.57,248.56) -- (176.94,290.06) .. controls (176.99,294.73) and (179.34,297.04) .. (184.01,297) ;

\draw (328.75,102.73) node [anchor=north west][inner sep=0.75pt]  [rotate=-45] [align=left] {future infinity};
\draw (209,109) node [anchor=north west][inner sep=0.75pt]   [align=left] {Phase II:};
\draw (215,130.4) node [anchor=north west][inner sep=0.75pt]    {$\ddot{a} \leq 0$};
\draw (217,230) node [anchor=north west][inner sep=0.75pt]   [align=left] {Phase I:};
\draw (223,251.4) node [anchor=north west][inner sep=0.75pt]    {$\ddot{a} \geq 0$};
\draw (67,233) node [anchor=north west][inner sep=0.75pt]   [align=left] {\textcolor[rgb]{0.82,0.01,0.11}{TCC-violating}};

\end{tikzpicture}
    \caption{The partial Penrose diagram of a flat expanding FRW spacetime with two phases. Phase I: contains epochs of accelerated expansion that collectively violate TCC. Phase II: an eternal decelerating phase or a finite decelerating phase followed by an eternal non-accelerating phase which, in either case lasts for infinite proper time in the comoving frame.}
    \label{Penrose1}
\end{figure}
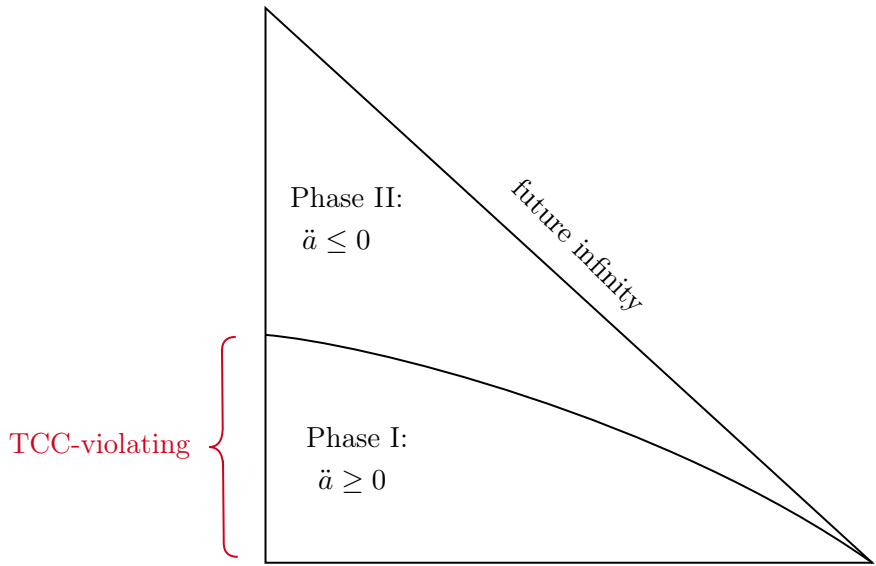

Let us first consider a homogeneous, isotropic, and spatially flat classical background that satisfies the above conditions. (We will later consider more general solutions.) Such a background has a spacetime  Penrose diagram that looks like Fig. \ref{Penrose1}. Phase I denotes the period of accelerated expansion which violates TCC and Phase II denotes the eternal phase that follows, which is not accelerating. Note that we are not concerned with the past infinity or whether the spacetime is geodesically complete in the past. What matters for our argument is that the semi-classical description is valid in the accelerating phase for long enough to violate the TCC. Therefore, we consider only a part of the spacetime that is bounded in the past.


Suppose a classical solution satisfying our conditions exists. 
The problem created by violating the TCC arises from the fact that the vacuum in Phase I is different from the vacuum in Phase II, and, consequently, a solution that begins in the vacuum of Phase I and transitions to Phase II necessarily is filled with particles and vice versa. In particular, this means that the vacuum of Phase II can be traced back in  Phase I to an excited state with trans-Planckian energies. Therefore, the energy-momentum tensor sourced by these trans-Planckian particles will cause a breakdown in the EFT. Note that the trans-Planckian particles are observable by the free-falling comoving observer. The existence of this solution means that this scalar potential cannot arise in the quantum gravity landscape under the assumption that the EFT must be valid in the future vacuum for any solution. As noted above, this assumption is motivated by holography and is a feature of how EFT is defined in perturbative string theory.

In our arguments, we use the deceleration in the future to infer the TCC in the past. One might find it surprising that we find a connection between the past and the future. This is indeed very surprising from the perspective of effective field theory, however, we would like to point out that in an expanding universe, the past corresponds to UV physics and the future corresponds to IR physics. Therefore, this is nothing other than the usual UV/IR connection in quantum gravity and holography.

The argument can be rephrased in terms of the contracting solution obtained by CPT conjugating the expanding solution that violates the TCC and satisfies our assumptions. CPT conjugation maps the future vacuum in the expanding solution to the past vacuum in the contracting solution. The contracting universe undergoes a transition from Phase II to Phase I which produces many particles with Hubble physical momentum. As the contraction continues in Phase I, these particles will get blue-shifted until they become trans-Planckian and break down the EFT.

How does the argument modify in the presence of anisotropy and spatial curvature? In the homogeneous limit in $d$-dimensional spacetimes,  anisotropy and spatial curvature contribute terms to the Friedmann equation that scale as $\propto 1/a^{2(d-1)}$ and $\propto 1/a^2$, respectively, to be compared to the energy density of the scalar field, which scales as $1/a^{\beta}$ where $\beta<2$ during Phase I and $2 \le \beta < 2(d-1)$ during Phase II.
Consequently, the ratio of the anisotropy to other contributions to the Friedmann equation is greatest at the beginning of Phase I and decreases monotonically and rapidly  during any expansion phases that follow.  Therefore, over the finite range of initial conditions in which the initial anisotropy  is negligible compared to the initial scalar field energy density at the beginning of Phase I,  the anisotropy will remain  negligible throughout the evolution and will not  violate any of the conditions assumed above.

\begin{figure}[h]
    \centering

\tikzset{every picture/.style={line width=0.75pt}} 

\begin{tikzpicture}[x=0.75pt,y=0.75pt,yscale=-0.9,xscale=0.9]

\draw   (316,10) -- (622,290) -- (316,290) -- cycle ;
\draw [color={rgb, 255:red, 65; green, 117; blue, 5 }  ,draw opacity=1 ][line width=1.5]    (316,175) .. controls (365,179) and (517,216) .. (622,290) ;
\draw  [color={rgb, 255:red, 208; green, 2; blue, 27 }  ,draw opacity=1 ] (301,176) .. controls (296.33,176.04) and (294.02,178.39) .. (294.06,183.06) -- (294.41,221.56) .. controls (294.47,228.23) and (292.17,231.58) .. (287.5,231.63) .. controls (292.17,231.58) and (294.53,234.89) .. (294.59,241.56)(294.57,238.56) -- (294.94,280.06) .. controls (294.99,284.73) and (297.34,287.04) .. (302.01,287) ;
\begin{scope}[shift={(12, 0)}]
\draw  [color={rgb, 255:red, 74; green, 144; blue, 226 }  ,draw opacity=1 ] (198,7) .. controls (193.33,7) and (191,9.33) .. (191,14) -- (191,64) .. controls (191,70.67) and (188.67,74) .. (184,74) .. controls (188.67,74) and (191,77.33) .. (191,84)(191,81) -- (191,159) .. controls (191,163.67) and (193.33,166) .. (198,166) ;
\end{scope}
\draw  [dash pattern={on 4.5pt off 4.5pt}]  (317,111) .. controls (392,114) and (556,235) .. (622,290) ;

\draw (426.75,78.73) node [anchor=north west][inner sep=0.75pt]  [rotate=-45] [align=left] {future infinity};
\draw (8,51) node [anchor=north west][inner sep=0.75pt]  [color={rgb, 255:red, 74; green, 144; blue, 226 }  ,opacity=1 ] [align=left] {Phase II: non-accelerating\\$\displaystyle \ddot{a} \leq 0$ eternal expansion};
\draw (12,221) node [anchor=north west][inner sep=0.75pt]  [color={rgb, 255:red, 208; green, 2; blue, 27 }  ,opacity=1 ] [align=left] {Phase I: has epochs of $\displaystyle \ddot{a} \geq 0$ that \\collectively violate TCC. };
\draw (222,134.4) node [anchor=north west][inner sep=0.75pt]    {$\Omega _{k} \ll 1$};
\draw (211,62) node [anchor=north west][inner sep=0.75pt]   [align=left] {\begin{minipage}[lt]{61.3pt}\setlength\topsep{0pt}
\begin{center}
$\displaystyle \Omega _{k}$: constant
\end{center}

\end{minipage}};
\draw (212,115) node [anchor=north west][inner sep=0.75pt]   [align=left] {Phase II-A:};
\draw (386,207.4) node [anchor=north west][inner sep=0.75pt]  [color={rgb, 255:red, 65; green, 117; blue, 5 }  ,opacity=1 ]  {$t=t_{*}$};
\draw (216,39) node [anchor=north west][inner sep=0.75pt]   [align=left] {Phase II-B:};
\draw (327,141.4) node [anchor=north west][inner sep=0.75pt]    {$\ddot{a} < 0$};
\draw (334,242.4) node [anchor=north west][inner sep=0.75pt]    {$\ddot{a}  >0$};
\draw (321,75.4) node [anchor=north west][inner sep=0.75pt]    {$aH< \infty $};

\end{tikzpicture}
    \caption{The Penrose diagram of a FRW solution with negative spatial curvature that transitions from a TCC-violating phase to a non-accelerating phase. Phase I has accelerating epochs that collectively violate the TCC. Phase II consists of two subphases: In Phase II-A,  $\ddot{a}$ is less than zero; the spatial curvature is negligible; and the expansion is entirely driven by the scalar field. In Phase II-B, $\ddot{a}$ becomes negligible;  the spatial curvature becomes important; and $\Omega_k$ converges to a non-zero value.  }
    \label{Penrose2}
\end{figure}
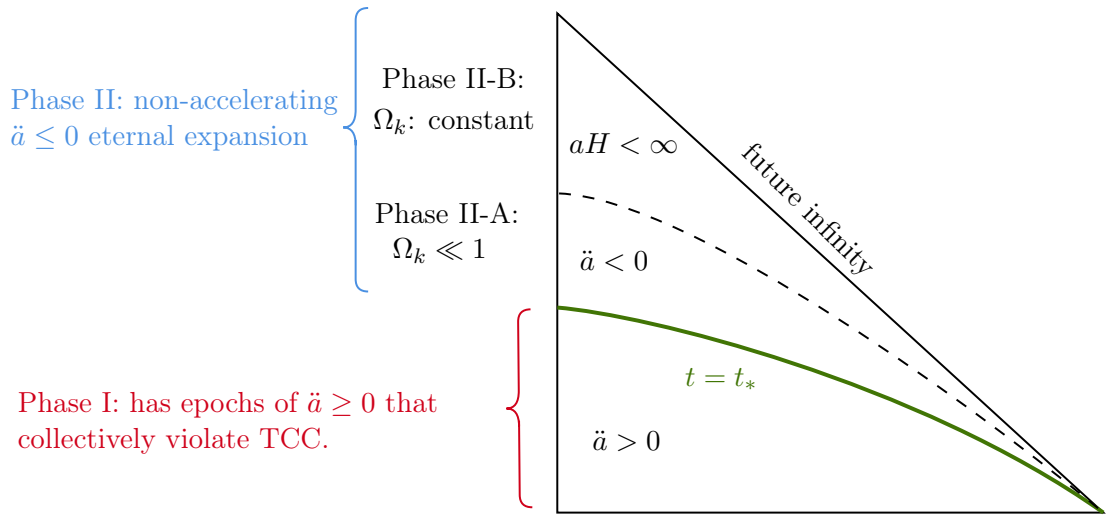

The story is more subtle for spatial curvature. Any non-zero amount of initial spatial curvature will eventually become non-negligible in Phase II after the accelerating Phase I has ended.  However, as we will show in Sec. \ref{Sec:curvature}, if there are spatially flat solutions that satisfy our assumptions, there is also a finite range of open universes that do as well. The key observation is that open universes can become non-accelerating ($\ddot a=0$) at future infinity. Moreover, there exists a finite range of initial spatial curvature for which the spatial curvature remains negligible until some  arbitrarily long (but finite) time after the accelerated expansion phase ends.  This range includes initial conditions for which our assumptions will still be satisfied. (Note that we do not consider closed universes, as they become singular in the future and hence violate assumption 2.)

\section{Particle production in cosmological transitions}\label{Sec:2}

In this section, for simplicity, we confine ourselves to flat FRW solutions that are described by classical equations of motion in which the evolution is driven by a scalar field $\phi$ with canonical energy density $-\frac{1}{2}\partial_\mu\phi\partial^\mu\phi$ and a positive scalar potential $V(\phi)$. The spacetime is described by the convential FRW metric
\begin{align}\label{metric}
    ds^2=-dt^2+a(t)^2\left(\sum_{i=1}^{d-1}(d x^i)^2\right)\,.
\end{align}
We examine the validity of EFT for  classical solutions satisfying conditions 1 thru 4  in Sec.~\ref{Sec:summary} that undergo a smooth transition between two phases of expansion: 
\begin{itemize}
\item  Phase I: a phase of accelerated expansion $\ddot{a}>0$ ($a(t)\sim t^q$, $q>1$) that lasts long enough to violate TCC; that is, $a_{\max}/a_{\min}>\Mpl/H_{\min}$ where $\max$ and $\min$ denote the maximum and minimum during this phase. In particular the accelerated phase can be a continuous solution consisting of a sequence of accelerating epochs interspersed with decelerating ones, as long as the epochs collectively violate the TCC. 
\item Phase II:  a phase of eternal decelerated expansion $\ddot{a} < 0$ ($a(t)\sim t^p$, $p<1$) that continues to the infinite future. 
\end{itemize}
The equation of state $w$ is given by $w = (2-3q)/3q$ (Phase I) and $w = (2-3p)/3p$ (Phase II) with $w = -1/3$ marking the separation between the two phases. In terms of the background behavior of the classical solution, this is similar to the usual inflationary cosmology. However, in inflationary cosmology, the conventional analysis assumes an accelerating Phase I whose {\it past vacuum} is Bunch-Davies \cite{Kaloper:2002cs}.  Here we are interested in examining classical solutions in which the decelerating Phase II terminates in a particle-free  {\it future vacuum}.   
 As we have discussed in the Sec.~\ref{Sec:summary}, our motivation stems from string theory and holography, which demand that EFT be valid for classical solutions bounded by the future vacuum.  

We will show in this section and Sec. \ref{Sec:curvature} that classical solutions bounded by the future vacuum state cannot be described by any EFT  if Phase I violates the TCC because such solutions will require that Phase I contains  excited states with physical trans-Planckian momentum. This gives a bottom up argument of why TCC-violation is incompatible with quantum gravity, given that the assumptions listed in Sec. \ref{Sec:summary} are satisfied.

\begin{figure}
    \centering

\tikzset{every picture/.style={line width=0.75pt}} 

\begin{tikzpicture}[x=0.75pt,y=0.75pt,yscale=-0.8,xscale=0.8]

\draw   (350,323) -- (653,41) -- (350,41) -- cycle ;
\draw    (351,175) .. controls (400,179) and (531,115) .. (653,41) ;
\draw [color={rgb, 255:red, 126; green, 211; blue, 33 }  ,draw opacity=1 ][line width=1.5]    (32,323) -- (32.99,25) ;
\draw [shift={(33,22)}, rotate = 90.19] [color={rgb, 255:red, 126; green, 211; blue, 33 }  ,draw opacity=1 ][line width=1.5]    (14.21,-4.28) .. controls (9.04,-1.82) and (4.3,-0.39) .. (0,0) .. controls (4.3,0.39) and (9.04,1.82) .. (14.21,4.28)   ;
\draw [color={rgb, 255:red, 208; green, 2; blue, 27 }  ,draw opacity=1 ]   (350,39.5) .. controls (351.67,37.83) and (353.33,37.83) .. (355,39.5) .. controls (356.67,41.17) and (358.33,41.17) .. (360,39.5) .. controls (361.67,37.83) and (363.33,37.83) .. (365,39.5) .. controls (366.67,41.17) and (368.33,41.17) .. (370,39.5) .. controls (371.67,37.83) and (373.33,37.83) .. (375,39.5) .. controls (376.67,41.17) and (378.33,41.17) .. (380,39.5) .. controls (381.67,37.83) and (383.33,37.83) .. (385,39.5) .. controls (386.67,41.17) and (388.33,41.17) .. (390,39.5) .. controls (391.67,37.83) and (393.33,37.83) .. (395,39.5) .. controls (396.67,41.17) and (398.33,41.17) .. (400,39.5) .. controls (401.67,37.83) and (403.33,37.83) .. (405,39.5) .. controls (406.67,41.17) and (408.33,41.17) .. (410,39.5) .. controls (411.67,37.83) and (413.33,37.83) .. (415,39.5) .. controls (416.67,41.17) and (418.33,41.17) .. (420,39.5) .. controls (421.67,37.83) and (423.33,37.83) .. (425,39.5) .. controls (426.67,41.17) and (428.33,41.17) .. (430,39.5) .. controls (431.67,37.83) and (433.33,37.83) .. (435,39.5) .. controls (436.67,41.17) and (438.33,41.17) .. (440,39.5) .. controls (441.67,37.83) and (443.33,37.83) .. (445,39.5) .. controls (446.67,41.17) and (448.33,41.17) .. (450,39.5) .. controls (451.67,37.83) and (453.33,37.83) .. (455,39.5) .. controls (456.67,41.17) and (458.33,41.17) .. (460,39.5) .. controls (461.67,37.83) and (463.33,37.83) .. (465,39.5) .. controls (466.67,41.17) and (468.33,41.17) .. (470,39.5) .. controls (471.67,37.83) and (473.33,37.83) .. (475,39.5) .. controls (476.67,41.17) and (478.33,41.17) .. (480,39.5) .. controls (481.67,37.83) and (483.33,37.83) .. (485,39.5) .. controls (486.67,41.17) and (488.33,41.17) .. (490,39.5) .. controls (491.67,37.83) and (493.33,37.83) .. (495,39.5) .. controls (496.67,41.17) and (498.33,41.17) .. (500,39.5) .. controls (501.67,37.83) and (503.33,37.83) .. (505,39.5) .. controls (506.67,41.17) and (508.33,41.17) .. (510,39.5) .. controls (511.67,37.83) and (513.33,37.83) .. (515,39.5) .. controls (516.67,41.17) and (518.33,41.17) .. (520,39.5) .. controls (521.67,37.83) and (523.33,37.83) .. (525,39.5) .. controls (526.67,41.17) and (528.33,41.17) .. (530,39.5) .. controls (531.67,37.83) and (533.33,37.83) .. (535,39.5) .. controls (536.67,41.17) and (538.33,41.17) .. (540,39.5) .. controls (541.67,37.83) and (543.33,37.83) .. (545,39.5) .. controls (546.67,41.17) and (548.33,41.17) .. (550,39.5) .. controls (551.67,37.83) and (553.33,37.83) .. (555,39.5) .. controls (556.67,41.17) and (558.33,41.17) .. (560,39.5) .. controls (561.67,37.83) and (563.33,37.83) .. (565,39.5) .. controls (566.67,41.17) and (568.33,41.17) .. (570,39.5) .. controls (571.67,37.83) and (573.33,37.83) .. (575,39.5) .. controls (576.67,41.17) and (578.33,41.17) .. (580,39.5) .. controls (581.67,37.83) and (583.33,37.83) .. (585,39.5) .. controls (586.67,41.17) and (588.33,41.17) .. (590,39.5) .. controls (591.67,37.83) and (593.33,37.83) .. (595,39.5) .. controls (596.67,41.17) and (598.33,41.17) .. (600,39.5) .. controls (601.67,37.83) and (603.33,37.83) .. (605,39.5) .. controls (606.67,41.17) and (608.33,41.17) .. (610,39.5) .. controls (611.67,37.83) and (613.33,37.83) .. (615,39.5) .. controls (616.67,41.17) and (618.33,41.17) .. (620,39.5) .. controls (621.67,37.83) and (623.33,37.83) .. (625,39.5) .. controls (626.67,41.17) and (628.33,41.17) .. (630,39.5) .. controls (631.67,37.83) and (633.33,37.83) .. (635,39.5) .. controls (636.67,41.17) and (638.33,41.17) .. (640,39.5) .. controls (641.67,37.83) and (643.33,37.83) .. (645,39.5) .. controls (646.67,41.17) and (648.33,41.17) .. (650,39.5) -- (653,39.5) -- (653,39.5)(350,42.5) .. controls (351.67,40.83) and (353.33,40.83) .. (355,42.5) .. controls (356.67,44.17) and (358.33,44.17) .. (360,42.5) .. controls (361.67,40.83) and (363.33,40.83) .. (365,42.5) .. controls (366.67,44.17) and (368.33,44.17) .. (370,42.5) .. controls (371.67,40.83) and (373.33,40.83) .. (375,42.5) .. controls (376.67,44.17) and (378.33,44.17) .. (380,42.5) .. controls (381.67,40.83) and (383.33,40.83) .. (385,42.5) .. controls (386.67,44.17) and (388.33,44.17) .. (390,42.5) .. controls (391.67,40.83) and (393.33,40.83) .. (395,42.5) .. controls (396.67,44.17) and (398.33,44.17) .. (400,42.5) .. controls (401.67,40.83) and (403.33,40.83) .. (405,42.5) .. controls (406.67,44.17) and (408.33,44.17) .. (410,42.5) .. controls (411.67,40.83) and (413.33,40.83) .. (415,42.5) .. controls (416.67,44.17) and (418.33,44.17) .. (420,42.5) .. controls (421.67,40.83) and (423.33,40.83) .. (425,42.5) .. controls (426.67,44.17) and (428.33,44.17) .. (430,42.5) .. controls (431.67,40.83) and (433.33,40.83) .. (435,42.5) .. controls (436.67,44.17) and (438.33,44.17) .. (440,42.5) .. controls (441.67,40.83) and (443.33,40.83) .. (445,42.5) .. controls (446.67,44.17) and (448.33,44.17) .. (450,42.5) .. controls (451.67,40.83) and (453.33,40.83) .. (455,42.5) .. controls (456.67,44.17) and (458.33,44.17) .. (460,42.5) .. controls (461.67,40.83) and (463.33,40.83) .. (465,42.5) .. controls (466.67,44.17) and (468.33,44.17) .. (470,42.5) .. controls (471.67,40.83) and (473.33,40.83) .. (475,42.5) .. controls (476.67,44.17) and (478.33,44.17) .. (480,42.5) .. controls (481.67,40.83) and (483.33,40.83) .. (485,42.5) .. controls (486.67,44.17) and (488.33,44.17) .. (490,42.5) .. controls (491.67,40.83) and (493.33,40.83) .. (495,42.5) .. controls (496.67,44.17) and (498.33,44.17) .. (500,42.5) .. controls (501.67,40.83) and (503.33,40.83) .. (505,42.5) .. controls (506.67,44.17) and (508.33,44.17) .. (510,42.5) .. controls (511.67,40.83) and (513.33,40.83) .. (515,42.5) .. controls (516.67,44.17) and (518.33,44.17) .. (520,42.5) .. controls (521.67,40.83) and (523.33,40.83) .. (525,42.5) .. controls (526.67,44.17) and (528.33,44.17) .. (530,42.5) .. controls (531.67,40.83) and (533.33,40.83) .. (535,42.5) .. controls (536.67,44.17) and (538.33,44.17) .. (540,42.5) .. controls (541.67,40.83) and (543.33,40.83) .. (545,42.5) .. controls (546.67,44.17) and (548.33,44.17) .. (550,42.5) .. controls (551.67,40.83) and (553.33,40.83) .. (555,42.5) .. controls (556.67,44.17) and (558.33,44.17) .. (560,42.5) .. controls (561.67,40.83) and (563.33,40.83) .. (565,42.5) .. controls (566.67,44.17) and (568.33,44.17) .. (570,42.5) .. controls (571.67,40.83) and (573.33,40.83) .. (575,42.5) .. controls (576.67,44.17) and (578.33,44.17) .. (580,42.5) .. controls (581.67,40.83) and (583.33,40.83) .. (585,42.5) .. controls (586.67,44.17) and (588.33,44.17) .. (590,42.5) .. controls (591.67,40.83) and (593.33,40.83) .. (595,42.5) .. controls (596.67,44.17) and (598.33,44.17) .. (600,42.5) .. controls (601.67,40.83) and (603.33,40.83) .. (605,42.5) .. controls (606.67,44.17) and (608.33,44.17) .. (610,42.5) .. controls (611.67,40.83) and (613.33,40.83) .. (615,42.5) .. controls (616.67,44.17) and (618.33,44.17) .. (620,42.5) .. controls (621.67,40.83) and (623.33,40.83) .. (625,42.5) .. controls (626.67,44.17) and (628.33,44.17) .. (630,42.5) .. controls (631.67,40.83) and (633.33,40.83) .. (635,42.5) .. controls (636.67,44.17) and (638.33,44.17) .. (640,42.5) .. controls (641.67,40.83) and (643.33,40.83) .. (645,42.5) .. controls (646.67,44.17) and (648.33,44.17) .. (650,42.5) -- (653,42.5) -- (653,42.5) ;
\draw    (343,175) -- (287,175) ;
\draw [shift={(285,175)}, rotate = 360] [color={rgb, 255:red, 0; green, 0; blue, 0 }  ][line width=0.75]    (10.93,-3.29) .. controls (6.95,-1.4) and (3.31,-0.3) .. (0,0) .. controls (3.31,0.3) and (6.95,1.4) .. (10.93,3.29)   ;
\draw    (361,41) -- (305,41) ;
\draw [shift={(303,41)}, rotate = 360] [color={rgb, 255:red, 0; green, 0; blue, 0 }  ][line width=0.75]    (10.93,-3.29) .. controls (6.95,-1.4) and (3.31,-0.3) .. (0,0) .. controls (3.31,0.3) and (6.95,1.4) .. (10.93,3.29)   ;

\draw (475.73,214.18) node [anchor=north west][inner sep=0.75pt]  [rotate=-315] [align=left] {past infinity};
\draw (368,186) node [anchor=north west][inner sep=0.75pt]  [color={rgb, 255:red, 74; green, 144; blue, 226 }  ,opacity=1 ] [align=left] {Phase II:};
\draw (374,87.4) node [anchor=north west][inner sep=0.75pt]  [color={rgb, 255:red, 208; green, 2; blue, 27 }  ,opacity=1 ]  {$\frac{d|\dot{a} |}{dt} < 0$};
\draw (376,60) node [anchor=north west][inner sep=0.75pt]  [color={rgb, 255:red, 208; green, 2; blue, 27 }  ,opacity=1 ] [align=left] {Phase I:};
\draw (9.5,197.5) node [anchor=north west][inner sep=0.75pt]  [color={rgb, 255:red, 126; green, 211; blue, 33 }  ,opacity=1 ,rotate=-270] [align=left] {time};
\draw (368,213.4) node [anchor=north west][inner sep=0.75pt]  [color={rgb, 255:red, 74; green, 144; blue, 226 }  ,opacity=1 ]  {$\frac{d|\dot{a} |}{dt} \geq 0$};
\draw (157,156) node [anchor=north west][inner sep=0.75pt]   [align=left] {\begin{minipage}[lt]{90.04pt}\setlength\topsep{0pt}
\begin{center}
Particle production \\with $\displaystyle k\sim H$
\end{center}

\end{minipage}};
\draw (176,284) node [anchor=north west][inner sep=0.75pt]   [align=left] {Initial vacuum};
\draw (65,284.4) node [anchor=north west][inner sep=0.75pt]    {$t\rightarrow -\infty :$};
\draw (71,161.4) node [anchor=north west][inner sep=0.75pt]    {$t=t_{*} :$};
\draw (64,32.4) node [anchor=north west][inner sep=0.75pt]    {$t=t_{*} +\Delta t:$};
\draw (156,30) node [anchor=north west][inner sep=0.75pt]   [align=left] {\begin{minipage}[lt]{97.4pt}\setlength\topsep{0pt}
\begin{center}
Blue-shifted particles\\are trans-Planckian
\end{center}

\end{minipage}};

\end{tikzpicture}
    \caption{The Penrose diagram of the CPT conjugated of the TCC-violating expanding solution that satisfies our assumptions. If we start with the past vacuum state, there will be particle production at the moment of transition from Phase II to Phase I. Under our assumptions, Phase I lasts long enough to blue-shift those particles to trans-Planckian energies that violate the EFT.}
    \label{Penrose3}
\end{figure}
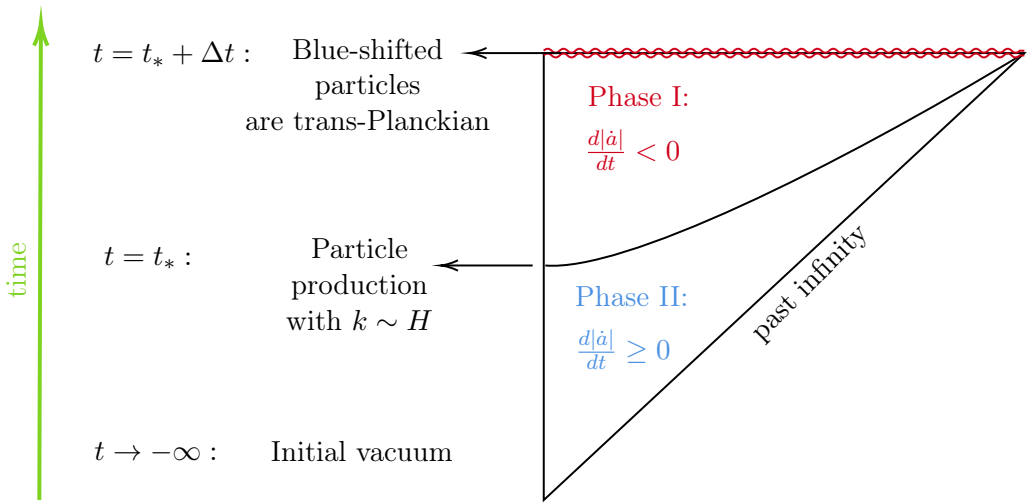

To demonstrate this conclusion, we can either 1) specify the ``final condition'' of the universe to be in the vacuum in Phase II and evolve it back in time to Phase I (as illustrated by the Penrose diagram in Fig.~\ref{Penrose1}); or, equivalently, 2) consider the CPT conjugate picture in which the universe begins in the particle-free future vacuum and is contracting, first through a period of decelerated contraction, Phase II, and, then, through a period of accelerated contraction, Phase I (as illustrated by the Penrose diagram in Fig.~\ref{Penrose3}). That is, here Phase II happens before Phase I.  Mathematically the two pictures are exactly equivalent and the same conclusions will be derived. We choose to work in the contracting picture such that all discussion of time evolution is going forward.

More specifically, in the contracting perspective, spacetime begins in a contracting Phase II with an infinite past and undergoes a period of decelerated contraction with $a(t)\sim (-t)^p,~p<1$, $\ddot{a}< 0$. This phase smoothly transitions into a contracting Phase I in which the spacetime undergoes a period of  accelerated contraction with $a(t)\sim (-t)^q,~q>1$, $\ddot{a}>0$. Note that the time coordinate $t$ is negative for this solution, $t\in(-\infty,t_f]$.  The label ``I'' and ``II'' refer to accelerating and decelerating phases respectively as before, but in the contracting perspective Phase II happens before Phase I in time and both phases are contracting. 

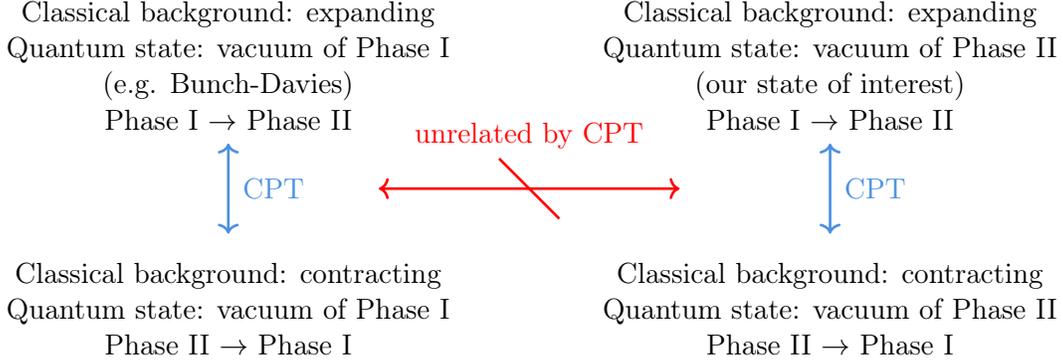
\begin{figure}
\centering
\begin{tikzpicture}[line width=1pt]
\node[align=center] at (0, 0) {Classical background: expanding\\
Quantum state: vacuum of Phase I\\
(e.g. Bunch-Davies)\\
Phase I $\rightarrow$ Phase II\\};

\draw[<->, color={rgb, 255:red, 74; green, 144; blue, 226 }] (0, -0.8)--(0, -2);
\node[align=center] at (0, -3) {Classical background: contracting\\
Quantum state: vacuum of Phase I\\
Phase II $\rightarrow$ Phase I};
\node[color={rgb, 255:red, 74; green, 144; blue, 226 }] at (0.6, -1.4) {CPT};

\begin{scope}[shift={(8, 0)}]
    \node[align=center] at (0, 0) {Classical background: expanding\\
Quantum state: vacuum of Phase II\\
(our state of interest)\\
Phase I $\rightarrow$ Phase II\\
};
\draw[<->, color={rgb, 255:red, 74; green, 144; blue, 226 }] (0, -0.8)--(0, -2);
\node[align=center] at (0, -3) {Classical background: contracting\\
Quantum state: vacuum of Phase II\\
Phase II $\rightarrow$ Phase I};
\node[color={rgb, 255:red, 74; green, 144; blue, 226 }] at (0.6, -1.4) {CPT};
\end{scope}

\draw[<->, color=red] (2, -1.4)--(6, -1.4);
\draw[red] (3.6, -1)--(4.4, -1.8);
\node at (4, -0.7) {\color{red}unrelated by CPT};
 
\end{tikzpicture}
\caption{CPT conjugate relation between different choice of states. The symbol $\rightarrow$ denote the order in time in which the two phases happen. Behavior of the universe in each CPT pair is completely equivalent. Our state of interest, which is a vacuum state in the decelerating Phase II, is not related by CPT to the usual Bunch-Davies setup in inflationary cosmology, which is a vacuum state in the accelerating Phase I.}
\label{fig:CPTpairs}
\end{figure}

Importantly, taking the CPT conjugate of our contracting solution beginning with the vacuum state of interest does \textbf{not} map it into an expanding solution in which the accelerating Phase I begin in the Bunch-Davies vacuum state.     
The relationship between the different choices of states and their CPT conjugates are illustrated in Fig. \ref{fig:CPTpairs}.

As described in Sec. \ref{Sec:summary}, the problem posed by TCC violation can be understood by considering a contracting solution starting with the vacuum state in Phase II and evolving in time to find the particle number of different modes in Phase I. As the vacuum state exits the horizon in Phase II and re-enters the horizon in Phase I, there is non-zero particle production at the moment of horizon re-entry. This mechanism of particle production seems similar to the usual inflation story. However, the key difference is that, unlike inflation, as we move away from the moment of transition, the excited modes are contracting and ultimately create trans-Planckian energy-densities that are inconsistent with an EFT.  

\subsection{A simple toy model (in the contracting picture)}\label{TM}

We consider a toy model that assumes an instantaneous transition between Phase II and Phase I. In this case $a(t)$ for the contracting solution is exactly known and all the calculations can be done exactly. We see that, right after the transition from Phase II to Phase I, there must be a large occupation number in the super-Hubble modes. If Phase I lasts long enough to violate the TCC, these highly occupied modes will blue-shift to  modes with trans-Planckian physical momentum. Therefore, the free-falling observer will see trans-Planckian particles and thus a breakdown of EFT. 

It is easier to work with the conformal time $a(\tau)d\tau=dt$. The metric in this coordinate takes the form
\begin{align}
    ds^2=a(\tau)^2\left[-d\tau^2+\sum_{i=1}^{d-1}(d x^i)^2\right]\,,
\end{align}
where prime will be used throughout to indicate derivatives with respect to conformal time $\tau$. In power-law FRW solutions, the conformal time has a minimum (maximum) for power-law contracting solutions that are eternally accelerating (decelerating). In our setup, the universe transitions between the two which makes $\tau$ unbounded from both below and above. However, we bound $\tau$ from above by considering only a finite period of Phase I that violates TCC.

In this toy model, the scale factor is given by 
\begin{align}\label{SFE}
        &\text{Phase II ($\tau<0$): }~~a(\tau)=\left(1-\frac{(1-p)H_0}{p}\tau\right)^\frac{p}{1-p}\,,\nonumber\\
        &\text{Phase I ($\tau>0$): }~~a(\tau)=\left(1+\frac{(q-1)H_0}{q}\tau\right)^{-\frac{q}{q-1}},
\end{align}
where $p<1$ and $q>1$ are positive constants. The scale factor is piece-wise defined, in other words $a(\tau)$ satisfies the usual $\dd t/a(t) = \dd \tau$ relation for $a(t)\sim (-t)^{p}$ and $a(t)\sim (-t)^{q}$ in each contracting phase respectively. The integration constants in the two contracting phases are chosen such that $a(\tau)$ and its first derivative are continuous at the instantaneous transition time $\tau=0$, and $a(\tau = 0) = 1$. $H_0$ is the magnitude of the Hubble parameter at the time of transition $\tau = 0$. In the frame of the comoving observer, this corresponds to a period of decelerating contraction that transitions into accelerated contraction instantaneously at $\tau = 0$. Note that $q=\infty$ corresponds to a de Sitter phase with exponential contraction in contracting Phase I.

The Hubble parameter $a'/a^2$ in this background is given by
\begin{align}\label{TMTE}
    &\text{Phase II ($\tau<0$): }~~H(\tau)=-H_0\left(1-\frac{(1-p)H_0}{p}\tau\right)^\frac{-1}{1-p},\nonumber\\
    &\text{Phase I ($\tau>0$): }~~H(\tau)=-H_0\left(1+\frac{(q-1)H_0}{q}\tau\right)^\frac{1}{q-1}.
\end{align}
Suppose the above cosmology is driven by a scalar field $\phi$ evolving along a positive potential $V(\phi)$. If the kinetic energy of the scalar field dominates the critical energy density, then $p=\frac{1}{d-1}$ where $d$ is the number of spacetime dimensions. Given that we are interested in the implications on the shape of the scalar potential, we only consider solutions in which  $p>\frac{1}{d-1}$ such that the scalar field potential is a non-negligible contribution in driving the contraction.

One can obtain power-law solutions for $a(t)$ from Eq. \eqref{TMTE} as attractor solutions for positive exponential potentials of the form
\begin{align}\label{SP}
    V(\phi)=\frac{d-2}{2\kappa^2}(d-1-\frac{1}{p})H_0^2\exp(-\kappa\frac{2}{\sqrt{(d-2)p}}\phi)\,,
\end{align}
where $\kappa=\sqrt{8\pi G}=\Mpl^{-(d-2)/2}$ is the reduced Planck factor, has a classical attractor solution 
\begin{align}
    \phi=\frac{\sqrt{(d-2)p}}{\kappa(1-p)}\ln(1-\frac{(1-p)H_0}{p}\tau)\,.
\end{align}
The scale factor for this attractor solution (setting $a(\tau=0)=1$) is given by 
\begin{align}
    a(\tau)=\left(1-\frac{(1-p)H_0}{p}\tau\right)^\frac{p}{1-p}.
\end{align}
Therefore, a potential with two exponential regimes that are glued together can approximate the time evolution in Eq. \eqref{TMTE}. However, to exactly reproduce this time evolution, $\phi'$ and $V(\phi)$ have to change discontinuously at $\tau=0$. 

In a realistic model with a continuous potential, the transition between the contracting phases will be smooth. We consider the glued-together time evolution only as a simple toy model that allows us to build intuition. In the next subsection, we will provide a general argument and as we will see, the discontinuity of the toy model does not play a significant role in the conclusions.

To define the vacuum, we need to determine the mode expansion for perturbations around the classical background solution. We review this procedure and our conventions in appendix \ref{A1}. The modes must satisfy the following two equations which are respectively the normalization condition and the equation of motion.
\begin{align}
&u(\tau,{\bf{k}}){u^*}'(\tau,{\bf{k}})-u'(\tau,-{\bf{k}})u^*(\tau,-{\bf{k}}))=ia(\tau)^{-(d-2)}\,\label{NC},\\
&u_k''(\tau)+(d-2)a(\tau)Hu_k'(\tau)+k^2u_k(\tau)+ a(\tau)^2\partial_\phi^2 V(\phi_0(\tau))u_k(\tau)\simeq 0\,\label{eom},
\end{align}
Due to the translational symmetry of the solution, we can consider the following quasi-stationary ansatz for $u$.
    \begin{align}
        u(\tau,k)=u_k(\tau)e^{i\vec k\cdot \vec X}\,.
    \end{align}
Moving forward, we drop the subindex and the argument of $u_k(\tau)$ and use $u$ for simplicity. After substituting in the scalar potential Eq. \eqref{SP} into the equation of motion Eq. \eqref{eom}, we find
\begin{align}
    -(d-2)\left[\frac{u'}{\frac{1}{H_0}-\frac{1-p}{p}\tau}\right]+u''+k^2u+\frac{H_0^2(p(d-1)-1)}{p^2(1-\frac{(1-p)H_0}{p}\tau)^2}u\simeq 0
\end{align}
Let us define 
\begin{align}
\tilde\tau=k(\tau-\frac{p}{(1-p)H_0})~~~\text{and}~~~U(\tilde\tau)=u(\tau)\,.
\end{align}
The equation of motion in terms of $U$ in contracting Phase II is 
    \begin{align}\label{UEOM}
       -(d-2)\frac{U'}{\tilde\tau}+(\frac{1-p}{p})[U''+U]+\frac{pd-p-1}{p(1-p)}\cdot\frac{U}{\tilde\tau^2}\simeq0,
    \end{align}
    Note that for $p=\infty$ which corresponds to the de Sitter space, the last term vanishes and we recover the usual equations in the de Sitter space.

The Eq. \eqref{NC} and Eq. \eqref{eom} do not uniquely determine the modes which is why the definition of the zero-particle state (vacuum) is not unique. To determine the modes, we must impose extra conditions that pick out the desired vacuum. We choose the modes in each contracting phase such that they asymptote to the Minkowski modes at large physical momentum.  

There is something very interesting about the Eq. \eqref{UEOM}. The behavior at  $\tau\rightarrow-\infty$ and fixed $k$ and the behavior at large $k$ and fixed $\tau$ are both captured by $\tilde\tau\rightarrow-\infty$. Therefore, if we make sure that that the mode has a single frequency at $\tilde\tau\rightarrow-\infty$, it would imply that the UV vacuum ($k\rightarrow\infty$) is identical to the early time vacuum ($\tau\rightarrow-\infty$). In other words, the state that has zero particles in the past infinity, also looks like the Minkowski vacuum at very short length scales and thus does not violate the equivalence principle. 
    
The differential Eq. \eqref{UEOM} has an exact solution in terms of  the Hankel functions:
   \begin{align}\label{RDM}
       U(\tilde\tau)=U_0(-\tilde\tau)^{\frac{1}{2}-\frac{p(d-2)}{2(1-p)}}H^{1}_{\nu(p)}(-\tilde\tau),
   \end{align}
    where 
    \begin{align}
        \nu(p)=\frac{\sqrt{\left((d-1)p-1\right)\left((d-1)p-5\right)}}{2|1-p|}.
    \end{align}
Hankel functions have the following asymptotic behavior.
    \begin{align}
        H_\nu^{(1,2)}(x)\xrightarrow{x\rightarrow\infty} \sqrt\frac{2}{\pi x}\exp(\mp i(x-\frac{\nu\pi}{2}-\frac{\pi}{4}))\,,
    \end{align}
and therefore their frequency at early times has a definite sign. The normalization condition Eq. \eqref{NC} in the limit $\tilde\tau\rightarrow-\infty$ gives
\begin{align}
    \frac{4}{\pi}|U_0|^2(-\tilde\tau)^{-\frac{p(d-2)}{1-p}}k=a(\tau)^{-(d-2)}=\left(-\frac{(1-p)H_0}{pk}\tilde\tau\right)^\frac{-p(d-2)}{1-p},
\end{align}
   which is solved by setting
   \begin{align}\label{MEC}
       U_0=\sqrt\frac{\pi}{4k}\left(\frac{(1-p)H_0}{pk}\right)^\frac{-p(d-2)}{2(1-p)}
   \end{align}
Using the Eq. \eqref{RDM} and \eqref{MEC} we find
\begin{align}
        \tau<0:~~u^{-,+}_k(\tau)=\sqrt{\frac{p\pi}{4H_0(1-p)}}\left(1-\frac{H_0(1-p)}{p}\tau\right)^{\frac{1}{2}-\frac{p(d-2)}{2(1-p)}}H^{1,2}_{\nu(p)}\left(k(\frac{p}{(1-p)H_0}-\tau)\right).
\end{align}
Repeating the same calculation for contracting Phase I leads to
\begin{align}
\tau>0:~~u^{-,+}_k(\tau)=\sqrt{\frac{q\pi}{4H_0(q-1)}}\left(1+\frac{(q-1)H_0}{q}\tau\right)^{\frac{1}{2}-\frac{q(d-2)}{2(1-q)}}H^{1,2}_{\nu(q)}\left(-k(\tau+\frac{q}{(q-1)H_0})\right)\,.
\end{align}

\begin{figure}[H]
    \centering
    \includegraphics[width=0.6\textwidth]{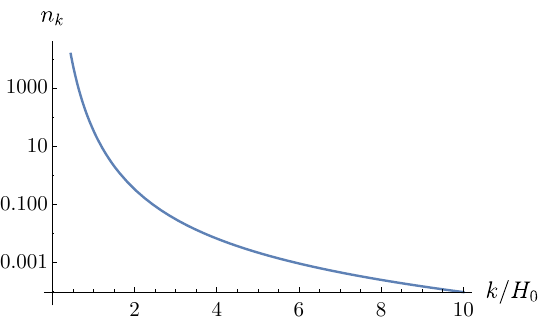}
    \caption{Particle number as a function of $k/H_0$, where $H_0$ is the  Hubble parameter at the time of transition $\tau = 0$, where we have normalized the scale factor such that $a(\tau = 0) = 1$. Here we choose $p = 1/2$ and $q = \infty$ as an example. There is a sharp increase in particle number for $k<H_0$, i.e. modes that are super-horizon at the time of transition. Similar sharp increase of $n_k$ at small $k$ is observed for all $p < 1$ and $q > 1$.}\label{fig:nk}
\end{figure}

By matching the zeroth and first order derivatives of the mode functions across $\tau=0$, we find the Bogoliubov coefficients $(\alpha_k,\beta_k)$ relating the two mode expansions.
    \begin{align}
u^+_k(0^-)&=\alpha_ku^+_k(0^+)+\beta_ku^-_k(0^+)\nonumber\\
u^{+'}_k(0^-)&=\alpha_ku^{+'}_k(0^+)+\beta_ku^{-'}_k(0^+).
    \end{align}
    which are given by
    \begin{align}
        \alpha_k&=\frac{u^{-'}_k(0^+)u^+_k(0^-)-u^-_k(0^+)u^{+'}_k(0^-)}{u^+_k(0^+)u^{-'}_k(0^+)-u^-_k(0^+)u^{+'}_k(0^+)}\nonumber\\
        \beta_k&=\frac{-u^{+'}_k(0^+)u^+_k(0^-)-u^+_k(0^+)u^{+'}_k(0^-)}{u^+_k(0^+)u^{-'}_k(0^+)-u^-_k(0^+)u^{+'}_k(0^+)}.
    \end{align}
The average particle number $n_k$ is given by 
\begin{align}
n_k=\beta_k\beta_k^*.
\end{align}
For all $q > 1$ and $p < 1$, there is a sharp increase in $n_k$ for all $k < H_0$. An example of  $n_k$ as a function of $k/H_0$ is shown in Fig. \ref{fig:nk}.

What is the physical interpretation of this increase in $n_k$ at $k < H_0$? It is perhaps helpful to draw a parallel between what we computed and the familiar inflationary mode function story. In the usual inflationary computation, we start with the Bunch-Davies vacuum state during the accelerating phase for all $k$ modes. Some of these $k$ modes exit the horizon during inflation and re-enter the horizon during the decelerating phase after inflation.  It is well known that particle production occurs when the mode re-enter the horizon in the decelerating phase. This process is often referred to as ``particle production in super horizon modes'', but it should be understood that actual particle production only occurs after horizon re-entry. When the modes are still super-horizon, there is not a well-defined vacuum state needed to define the particle number. Here the computation is exactly the time-reverse. We also start with the vacuum state in the initial phase (contracting Phase II) and evolve the classical contracting solution forward in time.  Modes in the vacuum state first ``exit'' the Hubble patch during contracting Phase II then ``re-enter'' the Hubble patch in the contracting Phase I.  We have found that there is strong particle production at this horizon ``re-entry''. In other words, we have shown that the initial vacuum state in contracting Phase II evolves into a highly excited state in contracting Phase I, with large particle numbers for all $k$ modes that exited the Hubble patch during contracting Phase II. By definition, these modes have $k/a(\tau) = |H|$ at the time of horizon exit. If we time-evolve these modes for $\Delta t \sim \frac{1}{H}\ln(\Mpl/H)$ in the contracting solution, they will have trans-Planckian physical momentum, causing EFT to break down. In other words, no EFT can describe the classical contracting solution that begins in the l vacuum state of Phase II  if the accelerating contraction in Phase I lasts longer than $\sim \frac{1}{|H|}\ln(\Mpl/|H|)$.

\begin{figure}
    \centering
    \includegraphics[width=0.5\textwidth]{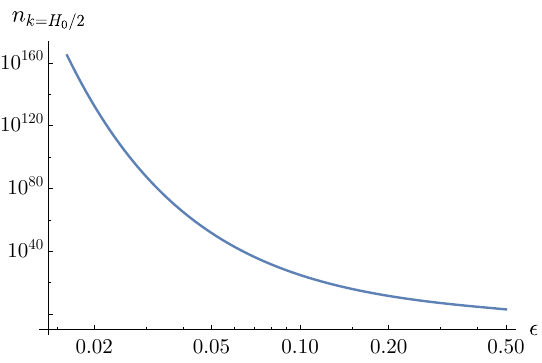}
    \caption{The particle number $n_k$ at $k = H_0/2$ for varying $\epsilon$, where we set $q = 1+\epsilon$ and $p = 1-\epsilon$. We see that as $\epsilon \rightarrow 0$, particle production at the time of transition diverges.}
    \label{fig:nk_eps}
\end{figure}
One might imagine that, if  the decelerating contraction phase and accelerating contraction phase were more similar in their contraction rate, i.e. $q=1+\epsilon$ and $p=1-\epsilon$ with $\epsilon\rightarrow 0$, there would be less particle production at the transition time $\tau = 0$.  Actually, the exact opposite happens. Fig. \ref{fig:nk_eps} shows the particle number at the transition time with $k = H_0/2$ as a function of $\epsilon$. As $\epsilon$ decreases, $n_k$ is divergent. In order to understand the discontinuity at $\epsilon=0$ recall that the vacuum of each phase is defined such that the sub-Hubble modes have zero-particles. Furthermore, the particle production is the consequence of exit and re-entry of the sub-Hubble modes.  In the $\epsilon\rightarrow0$ limit, the time required for this process diverges leading to infinite particle production. 

This might seem counterintuitive and in violation of the adiabiatic theorem which states that a sufficiently slow variation of the quantum system must evolve the initial vacuum into the final vacuum. However, the adiabatic theorem does not apply to quantum gravity since the Hamiltonian trivially vanishes and hence the vacuum cannot be defined as the energy ground state. In the $\epsilon\rightarrow 0$ limit, the vacua are increasingly different.

\subsection{WKB approximation (in the contracting picture)}\label{sec:WKB}
In this section, we use the Mukhanov-Sasaki equation and the WKB approximation to show that the discontinuity in $a''(\tau)$ in the toy model studied in the last section does not change our conclusion about the past vacuum corresponding to a highly excited state in the future. The Mukhanov-Sasaki equation in an arbitrary flat background driven by a scalar field is given by
\begin{equation}
    v_k''+\left(k^2-\frac{z''}{z}\right)v_k=0,
\end{equation}
where $v_k = a(\tau)u_k$, $z=a|\dot\phi/H| = 2 a^2 \epsilon$, $\epsilon = - \dot{H}/H^2 = 1/p$. We are interested in the solution to the Mukhanov-Sasaki equation in a general spacetime that in the past infinity has a power-law contraction $a\propto (-\tau)^\frac{p}{1-p}$ with $p< 1$ and in the future has a power-law contraction $a\propto \tau^\frac{q}{1-q}$ with $q>1$. In these two contracting phases we have
\begin{align}
    &\text{Phase II: } \frac{z''}{z}\simeq \frac{p(2p-1)}{(1-p)^2}\tau^{-2}\nonumber\\
    &\text{Phase I:  } \frac{z''}{z}\simeq \frac{q(2q-1)}{(1-q)^2}\tau^{-2}\,.
\end{align}
The case of de Sitter contraction in the future is recovered by $q\rightarrow \infty$, which gives $z''/z = 2 \tau^{-2}$. The Mukhanov-Sasaki equation takes the form
\begin{equation}
    v_k''-(\mathcal{V}(\tau)-k^2)v_k=0,
\end{equation}
where the \textit{WKB potential} function $\mathcal{V}(\tau)$ (not to be confused with the scalar potential) satisfies
\begin{align}
    &\lim_{\tau\rightarrow-\infty}\mathcal{V}(\tau)\tau^2\rightarrow \frac{p(2p-1)}{(1-p)^2}\,,\nonumber\\
    &\lim_{\tau\rightarrow \infty}\mathcal{V}(\tau)\tau^2\rightarrow \frac{q(2q-1)}{(1-q)^2}.\label{eq:Vtau_asymp}
\end{align}
We will assume there is a smooth transition between the two stages of expansion and $\mathcal{V}(\tau)$ has some finite maximum value near $\tau = 0$, which we denote by $\mathcal{V}_{\rm max}$.
\begin{figure}[h]  
\centering
\begin{tikzpicture}
\pgfplotsset{compat=newest, xmin=-8,xmax=8,ymin=-0.1,ymax=1.3}
\begin{axis}[
	width=8cm,
	height=5cm,
    axis x line=center,
    axis y line=center,
    xmajorticks=false,
    ymajorticks=false,
    xlabel={\large $\tau$},
    ylabel={\large $\mathcal{V}(\tau)$},
    clip mode = individual,
    every axis plot/.append style={thick}]
\addplot[mesh, draw=black, samples=100][domain=-7:-1.7]{3/x^2};
\addplot[mesh, draw=black, samples=100][domain=1:7]{1/x^2};
\node[black] at (axis cs:5,0.75){\large $A_k e^{- i k \tau}$};
\node[black] at (axis cs:5,0.5){\large $B_k e^{i k \tau}$};
\node[black] at (axis cs:-5,0.75){\large $e^{- i k \tau}$};
\draw [->, thick] (axis cs:3,0.75)--(axis cs:2,0.75);
\draw [<-, thick] (axis cs:3,0.5)--(axis cs:2,0.5);
	
\draw [->, thick] (axis cs:-2.5,0.75)--(axis cs:-3.5,0.75);
\end{axis}

\end{tikzpicture}
    \caption{Tunneling problem for calculating the future state obtained from the time evolution of the past positive frequency vacuum state.}
    \label{fig:WKB_tunneling}
\end{figure}
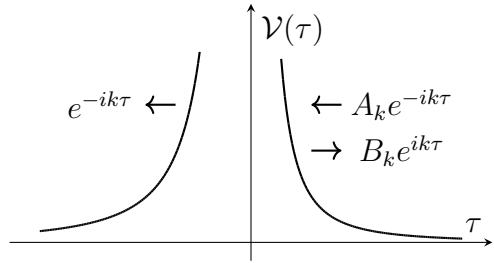

We want to find the behavior of the past positive frequency mode in terms of the future positive/negative frequency modes. Consider a solution with the following asymptotic behaviors,
\begin{align}
    &\tau\rightarrow +\infty: v_k\simeq A_ke^{-ik\tau}+B_ke^{ik\tau},\nonumber\\
    &\tau\rightarrow-\infty: v_k=e^{-ik\tau}.
\end{align}
For $k^2 < \mathcal{V}_{\rm max}$, this is mathematically equivalent to a ``tunneling'' problem as shown in Fig. \ref{fig:WKB_tunneling} (albeit with an unusual setup where we are tuning the amplitude of two incident waves on the left and the right such that there is no reflected wave on the left side of the potential.) The WKB approximation leads to $A_k$ and $B_k$ having exponentially enhanced amplitude with respect to $e^{-ik\tau}$ in the past,
\begin{align}\label{WKB}
    |A_k|=\sqrt{|B_k|^2+1}\sim\exp\left(\int_{\tau_1}^{\tau_2}\sqrt{\mathcal{V}(\tau)-k^2}\right),
\end{align}
where $\tau_1$ and $\tau_2$ are times at which $k^2=\mathcal{V}(\tau)$. When $k \ll \mathcal{V}_{\rm max}$, $\tau_1$ and $\tau_2$ occur in the regime where $\mathcal{V}_{\rm max}$ has the asymptotic form in Eq. \eqref{eq:Vtau_asymp}; therefore
\begin{align}
    &\tau_1\simeq -k^{-1}\sqrt{\frac{p(2p-1)}{(1-p)^2}},\nonumber\\
 &\tau_2\simeq k^{-1}\sqrt{\frac{q(2q-1)}{(1-q)^2}}.  
\end{align}
Note that the second expression is only correct for $p>\frac{1}{2}$. For $p<\frac{1}{2}$, $\mathcal{V}(\tau) \equiv z''/z$ is negative in the asymptotic future. In that case, the precise value of $\tau_1$ will then depend on the duration of the smooth transition between the accelerated and decelerated contraction. For simplicity we will assume $\tau_1 = 0$ for $p<\frac{1}{2}$. This approximation will not affect our main conclusions. Now let us use Eq. \eqref{WKB} to estimate the low-$k$ behavior of the coefficients $|A_k|$ and $|B_k|$.
\begin{align}
    k^2\ll \mathcal{V}_{\rm max}: |A_k|\simeq |B_k|&\sim \exp\left(\int^{k^{-1}\sqrt{\frac{q(2q-1)}{(1-q)^2}}}_{-k^{-1}\sqrt{\frac{p(2p-1)}{(1-p)^2}}}\sqrt{\mathcal{V}(\tau)-k^2}\right)\nonumber\\
    &\sim \exp\left(\sqrt\frac{p(2p-1)}{(1-p)^2}\ln(-\mathcal{V}_{\rm max}\tau_1)+\sqrt\frac{q(2q-1)}{(1-q)^2}\ln(\mathcal{V}_{\rm max}\tau_2)\right)\nonumber\\
    &\sim \left(\frac{k}{\mathcal{V}_{\rm max}}\right)^{-\sqrt\frac{q(2q-1)}{(1-q)^2}-\sqrt\frac{p(2p-1)}{(1-p)^2}}.
\end{align}
If $p<\frac{1}{2}$, the first term in the exponent disappears. Therefore, a more correct formula is 
\begin{align}
    k \ll \mathcal{V}_{\rm max}: |A_k|\simeq |B_k|\sim \left(\frac{k}{\mathcal{V}_{\rm max}}\right)^{-\sqrt\frac{q(2q-1)}{(1-q)^2}-\sqrt{\max\left(0,\frac{p(2p-1)}{(1-p)^2}\right)}}.\label{eq:AkBk_WKB1}
\end{align}
As one can see, at small momenta, there is always particle production. Moreover, note that if $p$ or $q$ are very close to one, the particle production is intensified, consistent with observation from the toy model in sec. \ref{TM}, as illustrated in Fig. \ref{fig:nk_eps}.

What exactly is the meaning of $k^2 <\max \mathcal{V}$ where we observe a strong particle production? Using the definition of $\mathcal{V}(\tau)$, we get
\begin{equation}
\begin{split}
\frac{k^2}{\Vc(\tau)} &= \frac{k^2 a}{a''} = \frac{k^2/a^2}{a''/a^3}\\
\frac{a''}{a^3} &= \left(\frac{\dot{a}}{a}\right)^2 + \frac{\ddot{a}}{a}= \dot{H} + 2H^2 =  \left(\frac{2q-1}{q}\right)H^2,
\end{split}
\end{equation}
where we have substituted the solution $a(t) \simeq a_0 (-t/t_0)^q$. So $k^2 <\mathcal{V}_{\rm max}$ means the comoving momentum $k^2$ is smaller than $\mathcal{V}_{\rm max} = \max \{a(\tau)^2 H(\tau)^2\left(\frac{2q-1}{q}\right)\}$. For $q> 1$, the comoving Hubble radius $(aH)^{-1}$ increases with time. Therefore $k^2 <\mathcal{V}_{\rm max}$ modes are the same modes that re-enter the Hubble patch in contracting Phase I (up to a factor of $(2q-1)/q \in [1, 2]$ for $q>1$). We have found particle production for all such $k$ modes. 

The WKB approximation supports the conclusion from Sec.~\ref{TM} that,  regardless of how the transition between decelerated and accelerated contraction phases happen, as long as the transition is smooth, the positive-frequency vacuum in the asymptotic past (Phase II) is described by a state with large particle number in small $k$ modes in the future (Phase I). These particle excitations have physical momentum $k/a = |H|$ at the time of horizon re-entry by definition, which means they would have trans-Planckian physical momentum $\log(\Mpl/|H|)$ number of e-folds after the transition. Therefore, if we have accelerated contraction in the future that is longer than $\log(\Mpl/|H|)$ number of e-folds, EFT cannot describe the vacuum state in the asymptotic past with a decelerated contraction.

\section{Adding spatial curvature}\label{Sec:curvature}

In the previous section we showed that the effective field theory breaks down for certain contracting solutions whose CPT conjugate violates TCC. However, the solutions that we considered were spatially-flat and isotropic. The space of classical solutions that are exactly isotropic and flat is ultra-fine-tuned and measure zero. Therefore, to make sure that our arguments are robust against small perturbations, we study the effect of small spatial curvature and anisotropy. In this section, we show that there is finite but non-zero range (i.e. measure) of open universe solutions for which our arguments continue to apply.

For any finite time interval of the solution, there is a non-zero range of sufficiently small perturbations that will not violate any of the assumptions listed in Sec. \ref{Sec:summary}. However, the same cannot be said about the asymptotic behavior of the solution. Assuming the contracting picture, the solutions that we had considered were bounded in the future (Phase I), but not bounded in the past (Phase II). Therefore, we have to worry about perturbations, that no matter how small, will alter the behavior of the solution at past infinity. 

The flat solutions that we considered in the previous section have the following asymptotic behavior at $t\rightarrow-\infty$. 
\begin{align}\label{SP2}
    &a(t)\sim (-t)^p~~(\frac{1}{d-1}<p<1)\nonumber\\
    & H^2\sim t^{-2}\sim a^{-2/p}\,.
\end{align}
We made the assumption $p>1/(d-1)$ to make sure the scalar potential energy remains non-negligible throughout the contraction. Any perturbation that has a contribution to the critical density which is strictly smaller than $H^2\sim a^{-2/p}$ in the past $t\rightarrow -\infty$  is ``safe''; that is, there exists a finite range of initial perturbations that remain negligible and do not violate any of our assumptions about the classical solution. For example, the contribution of anisotropy to energy density goes like $\propto a^{-2(d-1)}$ and therefore is safe because $p>1/(d-1)$. 

However, the spatial curvature contributes to the energy density as $\propto a^{-2}$ and always becomes relevant at past infinity. Thus, the addition of the spatial curvature can completely change the solution at past infinity. It is important to see how the solution changes and whether our arguments still apply to such modified spatially-curved solutions. 

It is instructive to study the effect of spatial curvature on the exponential potentials that generate spatially flat solutions such as (\ref{SP2}). Similar to the spatially-flat case, there is a tracking contracting solution which can be analytically computed. The FRW metric in the presence of spatial curvature is
\begin{align}\label{eq:FRWK}
    ds^2=-dt^2+a(t)^2\left[\frac{\dd r^2}{1-Kr^2}+ r^2 \dd \Omega^2\right]\, ,
\end{align}
where $K=1$ for a closed universe and $K=-1$ for an open universe. Suppose the scalar potential is given by 
\begin{align}
    V(\phi)=V_0\exp(-\kappa\lambda\phi)\,.
\end{align}
The Friedman equations in the presence of spatial curvature are 
\begin{align}
    &\frac{(d-1)(d-2)}{2}\left[H^2+\frac{K}{a^2}\right]=\frac{1}{2}\dot\phi^2+V(\phi)\,,\nonumber\\
    &\ddot\phi+(d-1)H\dot\phi+\frac{dV}{d\phi}=0\,.
\end{align}
The above equations have the following exact contracting solution.
\begin{align}\label{OUS}
    &a(t)=\frac{-t}{\sqrt{-K(1-\frac{4}{\lambda^2(d-2)})}}\nonumber\\
    &\phi(t)=\frac{\ln\left(\frac{\lambda^2 V_0}{2(d-2)}t^2\right)}{\lambda}\nonumber\\
    &\Omega_k=1-\frac{4}{\lambda^2(d-2)},~\Omega_{\dot\phi}=\frac{4}{\lambda^2(d-1)(d-2)},~\Omega_V=\frac{4}{\lambda^2(d-1)}\,
\end{align}
where the last line shows the fractional contribution  of the  spatial curvature ($\Omega_k$), kinetic energy density ($\Omega_{\dot\phi}$), and potential energy density ($\Omega_V$) to the Friedmann equation for the square of the Hubble parameter, $H^2$. Note that because of the square root in the expression for the scale factor, the tracking solution for negative spatial curvature only exists for $\lambda>\frac{2}{\sqrt{d-2}}$. This is the same range of $\lambda$ that we considered for the eternally decelerating solutions in the case of zero spatial curvature. So we focus on $\lambda>\frac{2}{\sqrt{d-2}}$ and $K=-1$. 

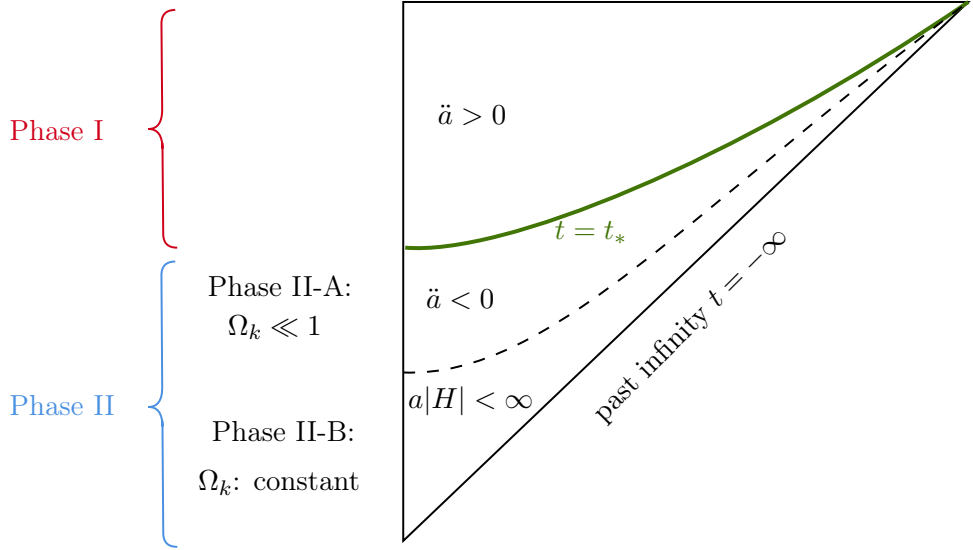
\begin{figure}[H]
    \centering
\tikzset{every picture/.style={line width=0.75pt}} 

\begin{tikzpicture}[x=0.75pt,y=0.75pt,yscale=-1,xscale=1]

\draw   (350,287) -- (635,15) -- (350,15) -- cycle ;
\draw [color={rgb, 255:red, 65; green, 117; blue, 5 }  ,draw opacity=1 ][line width=1.5]    (351,139) .. controls (400,143) and (515,97) .. (635,15) ;
\begin{scope}[shift={(-51, 0)}]
\draw  [color={rgb, 255:red, 208; green, 2; blue, 27 }  ,draw opacity=1 ] (286,19) .. controls (281.33,19.04) and (279.02,21.39) .. (279.06,26.06) -- (279.42,69.06) .. controls (279.47,75.73) and (277.17,79.08) .. (272.5,79.12) .. controls (277.17,79.08) and (279.53,82.39) .. (279.59,89.06)(279.56,86.06) -- (279.95,132.06) .. controls (279.99,136.73) and (282.34,139.04) .. (287.01,139) ;
\end{scope}
\draw  [color={rgb, 255:red, 74; green, 144; blue, 226 }  ,draw opacity=1 ] (235,146) .. controls (230.33,146.03) and (228.02,148.38) .. (228.05,153.05) -- (228.44,209.37) .. controls (228.49,216.04) and (226.18,219.39) .. (221.51,219.42) .. controls (226.18,219.39) and (228.53,222.7) .. (228.58,229.37)(228.56,226.37) -- (228.96,283.05) .. controls (228.99,287.72) and (231.34,290.03) .. (236.01,290) ;
\draw  [dash pattern={on 4.5pt off 4.5pt}]  (350,202) .. controls (425,205) and (497,117) .. (635,15) ;

\draw (441.73,220.25) node [anchor=north west][inner sep=0.75pt]  [rotate=-315] [align=left] {past infinity $\displaystyle t=-\infty $};
\draw (150,212) node [anchor=north west][inner sep=0.75pt]  [color={rgb, 255:red, 74; green, 144; blue, 226 }  ,opacity=1 ] [align=left] {Phase II};
\draw (150,73) node [anchor=north west][inner sep=0.75pt]  [color={rgb, 255:red, 208; green, 2; blue, 27 }  ,opacity=1 ] [align=left] {Phase I };
\draw (260,171.4) node [anchor=north west][inner sep=0.75pt]    {$\Omega _{k} \ll 1$};
\draw (245,250) node [anchor=north west][inner sep=0.75pt]   [align=left] {\begin{minipage}[lt]{61.3pt}\setlength\topsep{0pt}
\begin{center}
$\displaystyle \Omega _{k}$: constant
\end{center}

\end{minipage}};
\draw (250,152) node [anchor=north west][inner sep=0.75pt]   [align=left] {Phase II-A:};
\draw (425,124.4) node [anchor=north west][inner sep=0.75pt]  [color={rgb, 255:red, 65; green, 117; blue, 5 }  ,opacity=1 ]  {$t=t_{*}$};
\draw (252,225) node [anchor=north west][inner sep=0.75pt]   [align=left] {Phase II-B:};
\draw (360,158.4) node [anchor=north west][inner sep=0.75pt]    {$\ddot{a} < 0$};
\draw (366,65.4) node [anchor=north west][inner sep=0.75pt]    {$\ddot{a}  >0$};
\draw (351,207.4) node [anchor=north west][inner sep=0.75pt]    {$a|H|< \infty $};

\end{tikzpicture}
    \caption{The Penrose diagram of a contracting FRW solution with negative spatial curvature that undergoes to phase transitions. Phase I is a period of accelerating contraction long enough that shrinks Hubble modes to trans-Planckian wavelengths. Phase II consists of two subphases: In Phase II-B, $\ddot{a}$ becomes negligible;  the spatial curvature becomes important; and $\Omega_k$ converges to a non-zero value. In Phase II-A,  $\ddot{a}$ is less than zero; the spatial curvature is negligible; and the contraction is entirely driven by the scalar field.}
    \label{Penrose4}
\end{figure}

The contraction in Phase II can have two sub-phases, as illustrated in the Penrose diagram in Fig. \ref{Penrose4} which is the CPT conjugate of the expanding version illustrated in Fig. \ref{Penrose2}: 
\begin{itemize}
    \item Phase II-B in which $\Omega_k$ is non-negligible. This is the initial phase of contraction that starts at infite past.
    \item Phase II-A in which $\Omega_k$ is negligible and the solution is almost spatially flat. Contracting Phase II-A starts when the spatial-curvature $\propto a^{-2}\sim t^{-2p}$ becomes comparable to $H^2\sim 1/t^2$.
\end{itemize} 

Phase II-A is a strictly decelerating contraction, while Phase II-B could be marginally accelerating or decelerating based on how the tracking solution Eq. \eqref{OUS} transitions into contracting Phase II-A. For the argument in previous section to still be valid, we need to make sure that the addition of contracting Phase II-B does not violate the assumptions listed in Sec. \ref{Sec:summary}. In particular, we need all sub-Hubble modes in Phase I to trace back to sub-Hubble modes in the infinite past as required by assumption 3 in Sec. \ref{Sec:summary}. Suppose the mode that re-enters the horizon the latest in contracting Phase I has comoving momentum $k_{\min}$. Therefore, we would like $\frac{k_{\min}}{Ha}=\frac{k_{\min}}{\dot a}$ to be very small in the past infinity. Suppose $t_K$ and $t_*$ respectively denote the approximate times of the phase transitions Phase-IIB/Phase II-A and Phase II-A/Phase I.

\begin{align}\label{WT}
    \frac{k_{\min}}{aH}(t=-\infty)=\frac{k_{\min}}{\dot a}(t=-\infty)=\frac{\dot a(t_K)}{\dot a(t=-\infty)}\cdot \frac{\dot a(t_*)}{\dot a(t_K)}\cdot \frac{k_{\min}}{aH}(t_*),.
\end{align}
The decelerated contraction of Phase II-A ($a\sim (-t)^p$), lasts between times $t_K$ and $t_*$ which are both negative. This contracting phase contributes a factor of $\frac{\dot a(t_*)}{\dot a(t_K)}\sim \frac{-t_*}{-t_K}<1$ to the right hand side of Eq. \eqref{WT}. Therefore, by making the spatial curvature at $t_*$ smaller and prolonging Phase I, we can make the right hand side smaller. Note that the factor $\lim_{t\rightarrow-\infty}\frac{\dot a(t_K)}{\dot a(t)}$ always converges to a constant. However, depending on the details of the transition between Phase II-B to Phase IIA, that constant can be greater than one. 

Note that even if the relative ratio of physical wavelengths compared to the Hubble radius decreases in contracting Phase II-B, it would only by a finite factor. Therefore, by choosing a sufficiently small spatial curvature at $t=t_*$ and making $\frac{\dot a(t_*)}{\dot a(t_K)}$ sufficiently small, we can ensure that $\frac{k_{\min}}{a H}(t=-\infty)\ll 1$. In other words, we can always choose the spatial curvature to be sufficiently small (yet non-zero) that all the modes of interest  are  on wavelengths much smaller than the spatial curvature scale including throughout Phase IIB. Therefore, the mode analysis from Sec. \ref{sec:WKB} based on a spatially flat universe also applies for a range of sufficiently small spatial curvatures.

In the rest of this section we make this argument precise by showing that our WKB argument from Sec. \ref{sec:WKB} will still work if  we add a sufficiently small spatial curvature to the flat contracting solution at the time of transition between two contracting Phases II and I.

For the metric in Eq. \eqref{eq:FRWK} with non-zero spatial curvature,  the Mukhanov-Sasaki equation is~\cite{Garriga:1999vw}
\beq
v''_k + \left[k^2+2K\left(\frac{V'}{|\dot\phi H|}\right) - \frac{z''}{z}\right]v_k = 0,
\eeq
where
\beq
z = |\frac{a\dot\phi}{\mathcal{H}}|\sqrt{\frac{k^2 - 3 K }{k^2 + 2K\left(V'/|\dot\phi H|\right) }}\,.
\eeq

Note that $|V'|/\dot\phi H$ converges to a constant for the tracking solution at $\tau\rightarrow-\infty$. Therefore, assuming that it is a continuous function, it must have a maximum over $\tau\in[-\infty,\tau_f]$ for small spatial curvatures. Suppose this maximum is $k_{\max}^2$. For modes with $k\gg k_{\max}$, we can approximate $z\simeq |\frac{a\dot\phi}{\mathcal{H}}|$. This corresponds to the modes that have much shorter wavelengths than the length scale associated with the spatial curvature. The WKB potential is given by 
\begin{align}
    \mathcal{V}(\tau)=\frac{z''}{z}\,.
\end{align}
The relation between  conformal time $\tau$  and 
FRW time $t$ for the open universe solution given in Eq. \eqref{OUS} is
\begin{align}
    \tau =-\sqrt{1-\frac{4}{\lambda^2(d-2)}}\ln(-t)\,.
\end{align}
This implies that, if contraction in Phase II is an open universe, the WKB potential converges to the constant $\frac{1}{1-\frac{4}{\lambda^2(d-2)}}$ at past infinity $\tau\rightarrow-\infty$.

The WKB potential in contracting Phase II-B stays almost constant $\mathcal{V}(\tau)\simeq 1/(1-\frac{4}{\lambda^2(d-2)})$ while in contracting Phase II-A it grows as $\mathcal{V}(\tau)=z''/z\propto \tau^{-2}$.
We want all sub-horizon modes $k^2>\mathcal{V}$ at the end of Phase I ($\tau=\tau_f$) to trace back to sub-horizon modes in Phase II-B (i.e. $k^2>\mathcal{V}(\tau=-\infty)$).  This is equivalent to $\frac{\mathcal{V}(\tau_f)}{\mathcal{V}(-\infty)}\gg 1$. Let us see how $\frac{\mathcal{V}(\tau_f)}{\mathcal{V}(-\infty)}$ depends on the magnitude of the spatial curvature that we add to the flat solution in contracting Phase I. In the presence of spatial curvature and following the conventional normalization that  $K=-1$ for an open universe, the scale factor has dimensions and a physical meaning. 
Multiplying  the scale factor by $A>1$ at the transition between contracting Phases I and II-A is equivalent to decreasing the spatial curvature by a factor of $1/A^2$. Moreover, rescaling the scale factor, will also rescale the conformal time $d\tau=dt/a(t)$ as $\tau\rightarrow \tau/A$. Therefore, the WKB potential $\mathcal{V}=z''/z$ in Phase I will rescale as $\mathcal{V}_{\text{Phase I}}\rightarrow\mathcal{V}_{\text{Phase I}}A^2$. However, the asymptotic value of the WKB potential is fixed by the tracking solution to be $\mathcal{V}(-\infty)\simeq 1/(1-\frac{4}{\lambda^2(d-2)})$. Therefore, by decreasing the spatial curvature, we can arbitrarily increase the ratio $\frac{\mathcal{V}(\tau_f)}{\mathcal{V}(-\infty)}$ until all of the physical modes that reenter the Hubble horizon in Phase I undergo particle production.

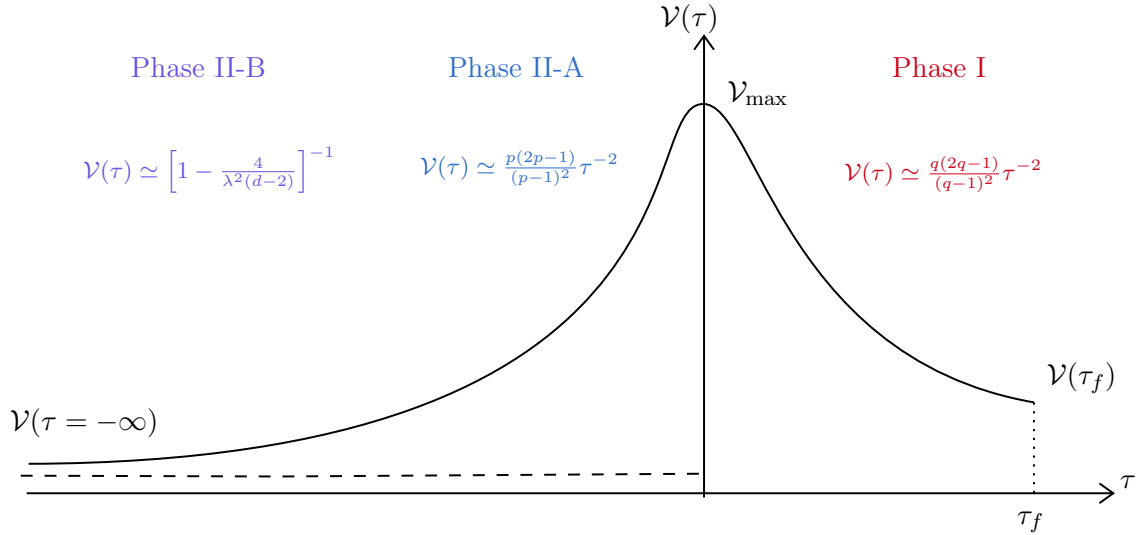
\begin{figure}[H]
    \centering

\tikzset{every picture/.style={line width=0.75pt}} 

\begin{tikzpicture}[x=0.75pt,y=0.75pt,yscale=-1,xscale=1]

\draw  (33,261.87) -- (581,261.87)(374.61,31) -- (374.61,264) (574,256.87) -- (581,261.87) -- (574,266.87) (369.61,38) -- (374.61,31) -- (379.61,38)  ;
\draw    (34,247) .. controls (374,246) and (342,77) .. (371,66) .. controls (400,55) and (409,193) .. (541,216) ;
\draw  [dash pattern={on 4.5pt off 4.5pt}]  (30,253) -- (161,253.4) -- (374,252) ;
\draw  [dash pattern={on 0.84pt off 2.51pt}]  (541,216) -- (541,262) ;

\draw (351,13.4) node [anchor=north west][inner sep=0.75pt]    {$\mathcal{V} ( \tau )$};
\draw (468,40) node [anchor=north west][inner sep=0.75pt]  [color={rgb, 255:red, 208; green, 2; blue, 27 }  ,opacity=1 ] [align=left] {Phase I};
\draw (244,40) node [anchor=north west][inner sep=0.75pt]  [color={rgb, 255:red, 46; green, 110; blue, 202 }  ,opacity=1 ] [align=left] {Phase II-A};
\draw (60,85.4) node [anchor=north west][inner sep=0.75pt]  [font=\footnotesize,color={rgb, 255:red, 107; green, 80; blue, 227 }  ,opacity=1 ]  {$\mathcal{V} ( \tau ) \simeq \left[ 1-\frac{4}{\lambda ^{2}( d-2)}\right]^{-1}$};
\draw (444,90.4) node [anchor=north west][inner sep=0.75pt]  [font=\footnotesize,color={rgb, 255:red, 208; green, 2; blue, 27 }  ,opacity=1 ]  {$\mathcal{V} ( \tau ) \simeq \frac{q( 2q-1)}{( q-1)^{2}} \tau ^{-2}$};
\draw (385,52.4) node [anchor=north west][inner sep=0.75pt]    {$\mathcal{V} _{\max}$};
\draw (230,87.4) node [anchor=north west][inner sep=0.75pt]  [font=\footnotesize,color={rgb, 255:red, 46; green, 110; blue, 202 }  ,opacity=1 ]  {$\mathcal{V} ( \tau ) \simeq \frac{p( 2p-1)}{( p-1)^{2}} \tau ^{-2}$};
\draw (84,40) node [anchor=north west][inner sep=0.75pt]  [color={rgb, 255:red, 107; green, 80; blue, 227 }  ,opacity=1 ] [align=left] {Phase II-B};
\draw (23,217.4) node [anchor=north west][inner sep=0.75pt]    {$\mathcal{V} ( \tau =-\infty )$};
\draw (583,251.4) node [anchor=north west][inner sep=0.75pt]    {$\tau $};
\draw (531,269.4) node [anchor=north west][inner sep=0.75pt]    {$\tau _{f}$};
\draw (546,194.4) node [anchor=north west][inner sep=0.75pt]    {$\mathcal{V} ( \tau _{f})$};

\end{tikzpicture}
\label{WKBP}
\caption{The WKB potential $\mathcal{V}=z''/z$ versus the conformal time. Phase II-B with non-negligible spatial curvature with unbounded past. Phase II-A is a period of deccelerated contraction with negligible spatial curvature. Phase I denotes a finite period of accelerated contraction which lasts long enough to violate TCC. By choosing a sufficiently small spatial curvature at $\tau=0$, we can make $\mathcal{V}(\tau_f)/\mathcal{V}(-\infty)$ arbitrarily large.}
\end{figure}

\section{Critical points of the scalar potential}\label{Sec:3}

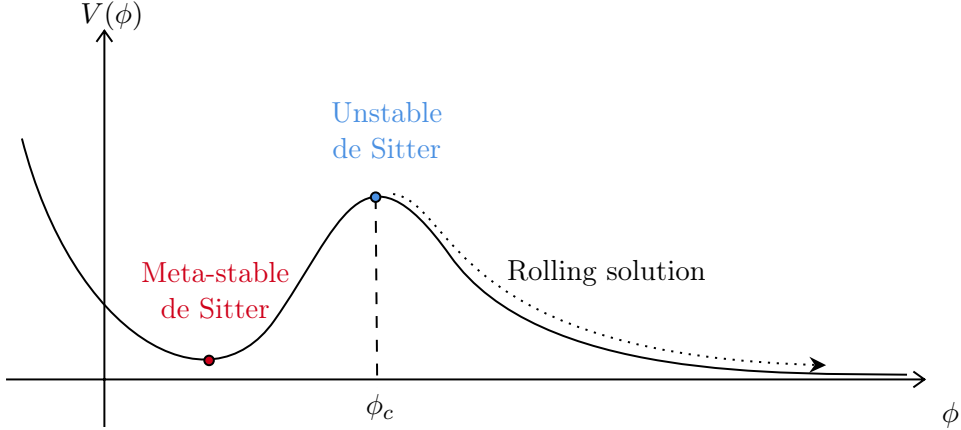
\begin{figure}[H]
    \centering

\tikzset{every picture/.style={line width=0.75pt}} 

\begin{tikzpicture}[x=0.75pt,y=0.75pt,yscale=-0.8,xscale=0.8]

\draw  [dash pattern={on 4.5pt off 4.5pt}]  (255,156) -- (256,276) ;
\draw  (22,271) -- (601,271)(84,51) -- (84,301) (594,266) -- (601,271) -- (594,276) (79,58) -- (84,51) -- (89,58)  ;
\draw    (32,119) .. controls (64,244) and (150,292) .. (191,234) .. controls (232,176) and (246,114) .. (302,193) .. controls (358,272) and (512,267) .. (590,268) ;
\draw  [fill={rgb, 255:red, 74; green, 144; blue, 226 }  ,fill opacity=1 ] (252,156) .. controls (252,154.34) and (253.34,153) .. (255,153) .. controls (256.66,153) and (258,154.34) .. (258,156) .. controls (258,157.66) and (256.66,159) .. (255,159) .. controls (253.34,159) and (252,157.66) .. (252,156) -- cycle ;
\draw  [fill={rgb, 255:red, 208; green, 2; blue, 27 }  ,fill opacity=1 ] (147,259) .. controls (147,257.34) and (148.34,256) .. (150,256) .. controls (151.66,256) and (153,257.34) .. (153,259) .. controls (153,260.66) and (151.66,262) .. (150,262) .. controls (148.34,262) and (147,260.66) .. (147,259) -- cycle ;
\draw  [dash pattern={on 0.84pt off 2.51pt}]  (266,154) .. controls (305,160) and (300,257) .. (539,262) ;
\draw [shift={(539,262)}, rotate = 181.2] [fill={rgb, 255:red, 0; green, 0; blue, 0 }  ][line width=0.08]  [draw opacity=0] (10.72,-5.15) -- (0,0) -- (10.72,5.15) -- (7.12,0) -- cycle    ;

\draw (67,30.4) node [anchor=north west][inner sep=0.75pt]    {$V( \phi )$};
\draw (610,283.4) node [anchor=north west][inner sep=0.75pt]    {$\phi $};
\draw (247,278.4) node [anchor=north west][inner sep=0.75pt]    {$\phi _{c}$};
\draw (222,94) node [anchor=north west][inner sep=0.75pt]  [color={rgb, 255:red, 74; green, 144; blue, 226 }  ,opacity=1 ] [align=left] {\begin{minipage}[lt]{45.82pt}\setlength\topsep{0pt}
\begin{center}
Unstable \\de Sitter
\end{center}

\end{minipage}};
\draw (103,195) node [anchor=north west][inner sep=0.75pt]  [color={rgb, 255:red, 208; green, 2; blue, 27 }  ,opacity=1 ] [align=left] {\begin{minipage}[lt]{58.85pt}\setlength\topsep{0pt}
\begin{center}
Meta-stable \\de Sitter
\end{center}

\end{minipage}};
\draw (336,194) node [anchor=north west][inner sep=0.75pt]   [align=left] {Rolling solution};
\end{tikzpicture}
\caption{A scalar potential with two critical points corresponding to 1) a metastable de Sitter vacuum in red, and 2) an unstable de Sitter in blue. The dotted curve shows the trajectory of a classical solution that starts very close to the unstable de Sitter and after a quasi-de Sitter phase transitions into a non-accelerating solution by rolling down the potential.}\label{DSCP}
\end{figure}

The arguments of Sec. \ref{Sec:2} apply to any classical solution that undergoes a transition from TCC-violating accelerating expansion to an eternal decelerating expansion. If the potential has a critical point $\partial_\phi V=0$ where the potential is positive $V>0$, one way to generate the accelerating phase is to consider a solution in which the scalar field is very close to that critical point and has zero kinetic energy. If the scalar field is \textit{exactly} at the critical point, it will stay there and the resulting classical solution is a de Sitter spacetime. However, the stability of the solution depends on the convexity of $V$ at the critical point. For $\partial_\phi^2V>0$, the solution is a minimum and classically stable, while for $\partial_\phi^2V<0$, perturbations of the scalar field away from the critical point will grow (Fig. \ref{DSCP}). The classically stable solution can still be quantum mechanically unstable via quantum tunneling to a more stable vacuum. Since minima with $V>0$ 
are generally believed to be local but not global minima in string theory, we shall refer to such solutions as metastable de Sitter vacua.

The de Sitter solutions that are classically unstable provide a good setup for our argument in Sec. \ref{Sec:2}. For example, consider a scalar field potential $V(\phi)$ which has a local maximum at $\phi=\phi_c$ and has a steep exponential decay for $\phi>\phi_c$ (Fig. \ref{DSCP}). By choosing the initial condition to be $(\phi,\dot\phi)(t=0)=(\phi_c+\epsilon,0)$ for sufficiently small $\epsilon$, we can make the quasi-de Sitter accelerating phase arbitrarily long so that it violates TCC. However, eventually the scalar field will roll down the potential and the solution will transition into a decelerating background. Therefore, our argument applies to this setup and would rule out such classical solutions. 

A scalar potential can still have an unstable critical point as long as quantum effects make it sufficiently short-lived. The quantum fluctuations of the scalar field about the local maximum will exponentially grow with a rate of $\tau_\text{quantum}^{-1}$. For $\tau<\tau_\text{quantum}$, the background will remain classical and we can apply our argument. Suppose $H$ is the Hubble parameter for this quasi-de Sitter state. If $\tau_\text{quantum}>\frac{1}{H}\ln(\Mpl/H)$, there can exist initial conditions in which the solution remains accelerating for $\tau$ that is less than $\tau_\text{quantum}$ but longer than $\frac{1}{H}\ln(\Mpl/H)$. That is, there exists in this case a solution that remains classical but violates TCC, where we can apply our argument and rule out the solution. Therefore, consistency with quantum gravity requires that the theory satisfy the condition
\begin{align}\label{UTCC}
    \tau_\text{quantum}\lesssim \frac{1}{H}\ln(\Mpl/H),
\end{align}
which is exactly the statement of TCC for local maxima of the effective potential~\cite{Bedroya:2019snp}. In particular, this implies that the second derivative of the potential must be large enough such that quantum fluctuations can quickly drive the state away from the local maximum.  Since the evolution of the quantum fluctuations depends on the shape of the potential near the critical point, the exact condition takes a complicated form for general potentials (see equation B.1 in~\cite{Bedroya:2019snp}). 

For simplicity, consider an inverted quadratic potential $V=V_0[1-\frac{\eta}{2}(\phi-\phi_c)^2]$ in four dimensions defined over a field range $\phi\in[\phi_c,\phi_c+\Delta\phi]$. We assume the potential is sufficiently small in Planck units ($V_0/(\eta(\Delta\phi)^2)\ll M_{\rm pl}^4$). Then the condition Eq. \eqref{UTCC} leads to
\begin{align}
    \eta\geq \frac{8V(\phi_c+\Delta\phi)}{3V_0}\left[\ln\left(\frac{3M_{\rm pl}^4}{V(\phi_c+\Delta\phi)}\right)\right]^{-2}\,.
\end{align}

We can easily generalize the above argument to the multi-field scalar potentials. Suppose the scalar potential depends on multiple scalar fields $\phi^I$ with a kinetic term $\frac{G_{IJ}(\vec \phi)}{2}\partial_\mu\phi^I\partial_\nu\phi^J$ and the potential has a critical point $\phi=\phi_c$ at which $\partial V/\partial\phi^I=0$ for every $I$. The kinetic term of the scalar fields impose a canonical metric $G_{IJ}$ on the space of scalar fields. Using this metric, we choose an normal basis $\delta \phi'^I$ for the tangent space of the scalar field space at point $\phi_c$. In this basis, we have $G_{IJ}=\delta_{IJ}$ and $\partial_{\delta\phi'^I} G_{JK}=0$. The critical point $\phi_c$ is an unstable saddle if the smallest eigenvalue of the Hessian matrix $\partial_{\delta\phi'^I}\partial_{\delta\phi'^J} V$ is negative. Let us denote the most negative eigenvalue as $-\eta_{\max}V(\phi_c)<0$ where $\eta_{\max}$ is a positive number. Therefore, there is a direction in the scalar field space in which the scalar potential is locally approximated by $V\simeq V(\phi_c)[1-\frac{\eta_{\max}}{2}(\Delta\phi)^2]$ where $\Delta\phi$ is the distance from the critical point $\phi_c$ measured by the canonical metric $G_{IJ}$. If we can find a solution that moves away from the unstable equilibrium at $\phi_c$  and transitions into a decelerated expansion, we can apply our single-field analysis noted above to find the quantum gravity constraint
\begin{align}
    \eta_{\max}\gtrsim \left[\ln(\frac{M_{\rm pl}^d}{V(\phi_c)})\right]^{-2}\,.
\end{align}
Just as in the single-field case, the exact inequality will depend on the details of the scalar potential.

In the above discussion we focused on unstable de Sitter solutions rather than meta-stable solutions given by the local minima of the scalar potential. This is because assumption 4 from Sec. \ref{Sec:summary} requires a transition between the accelerating phase and the decelerating phase that is well-described by the classical equations of motion. A metastable de Sitter corresponding to the local minimum of the potential can only evolve into an eternally decelerating background through non-perturbative quantum tunneling, which violates this assumption. Even if we start with a bubble of the meta-stable vacuum in a decelerating background, the assumptions will not be automatically satisfied since the bubble typically does not disappear \cite{Blau:1986cw}. However, for a scalar potential with a positive local minimum, we can consider other solutions by varying the initial kinetic energy of the scalar field. If any such solution satisfies the assumptions listed in Sec. \ref{Sec:summary} and violates TCC, the scalar potential can be ruled out. 

\begin{figure}[H]
    \centering

\tikzset{every picture/.style={line width=0.75pt}} 

\begin{tikzpicture}[x=0.75pt,y=0.75pt,yscale=-0.8,xscale=0.8]

\draw  (25,301.44) -- (646,301.44)(91.5,26) -- (91.5,339) (639,296.44) -- (646,301.44) -- (639,306.44) (86.5,33) -- (91.5,26) -- (96.5,33)  ;
\draw    (16,83) .. controls (25,109) and (71,192) .. (144,197) .. controls (217,202) and (286,159) .. (347,231) .. controls (408,303) and (553,295) .. (631,296) ;
\draw  [fill={rgb, 255:red, 208; green, 2; blue, 27 }  ,fill opacity=1 ] (151,197) .. controls (151,195.34) and (152.34,194) .. (154,194) .. controls (155.66,194) and (157,195.34) .. (157,197) .. controls (157,198.66) and (155.66,200) .. (154,200) .. controls (152.34,200) and (151,198.66) .. (151,197) -- cycle ;
\draw  [dash pattern={on 4.5pt off 4.5pt}]  (97,171) .. controls (147.49,206.64) and (246.99,157.01) .. (315.92,189.98) ;
\draw [shift={(318,191)}, rotate = 206.9] [fill={rgb, 255:red, 0; green, 0; blue, 0 }  ][line width=0.08]  [draw opacity=0] (10.72,-5.15) -- (0,0) -- (10.72,5.15) -- (7.12,0) -- cycle    ;

\draw (45,15.4) node [anchor=north west][inner sep=0.75pt]    {$V( \phi )$};
\draw (643,317.4) node [anchor=north west][inner sep=0.75pt]    {$\phi $};
\draw (108,215) node [anchor=north west][inner sep=0.75pt]  [color={rgb, 255:red, 208; green, 2; blue, 27 }  ,opacity=1 ] [align=left] {\begin{minipage}[lt]{82.66pt}\setlength\topsep{0pt}
\begin{center}
Wide and shallow\\local minimum
\end{center}

\end{minipage}};
\draw (238,124) node [anchor=north west][inner sep=0.75pt]   [align=left] {\begin{minipage}[lt]{73.58pt}\setlength\topsep{0pt}
\begin{center}
Quasi-de Sitter \\solution
\end{center}

\end{minipage}};

\end{tikzpicture}
    \caption{If a scalar potential has a positive local minimum which is sufficiently wide and shallow, there is TCC-violating classical solution that rolls out of the dip.}
    \label{MSDS}
\end{figure}
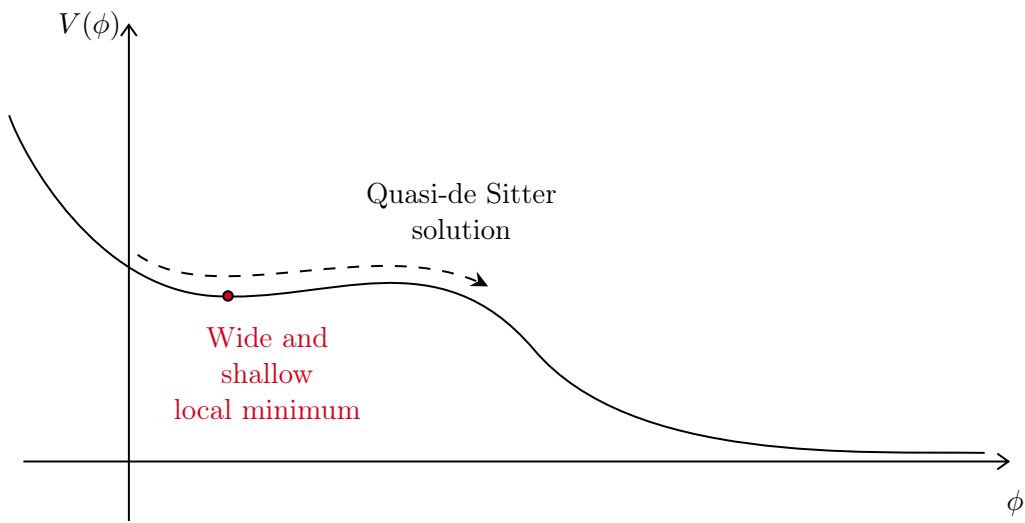

Let us consider an example of how, by choosing the proper initial condition, some potentials with positive local minima can be ruled out. Consider a positive potential with a very shallow dip such that $|V'|/V$ remains very small between the local minimum and the local maximum and the barrier height is much smaller than the value of the potential (see Fig. \ref{MSDS}). For such a potential, a classical solution in the local minimum can overcome the barrier and roll into a decelerating solution with a very small kinetic energy. Therefore, such a potential supports a solution that satisfies our assumptions. If the local minimum is sufficiently wide, the quasi-de Sitter portion of the solution can last long enough to violate the TCC and the potential will be ruled out. In the presence of more than one scalar field, there is even more freedom to find TCC-violating solutions that classically roll outside of the local minimum. In order to see exactly what potentials are consistent with our constraints, one has to consider all classical trajectories and check if TCC is violated for any solution that satisfies the assumptions listed in Sec. \ref{Sec:summary}.


\section{Trans-Planckian problem in black hole physics vs cosmology}\label{sec:black hole}

In this section, we explain the difference between our argument and ``the trans-Planckian problem'' raised decades ago in inflationary cosmology and black hole physics \cite{Martin:2000xs,Brandenberger:2012aj,Jacobson:1991gr} and why our argument applies to cosmological setups but does not apply to black holes.

The trans-Planckian problem questioned the validity of EFT when a quasi-stationary mode has a sub-Planckian wavelength in part of the spacetime. For example, in expanding cosmologies, the quasi-stationary modes can be trans-Planckian in the past. Similarly, the stationary modes outside of a Schwarzschild black hole become trans-Planckian near the black hole horizon due to the gravitational blue-shift. However, there is overwhelming evidence that the EFT calculation in black hole backgrounds that result in Hawking radiation must be trusted. 

The argument presented in this paper is different, not simply a reiteration of the trans-Planckian problem discussed previously. The trans-Planckian problem in cosmology is typically addressed by restricting the EFT Hilbert space to states for which the trans-Planckian states are in the vacuum \cite{Kaloper:2002cs}. Crucially, we consider a different quantum state. Based on holography and perturbative string theory, we argue that the vacuum of the asymptotic future boundary of spacetime must also admit an EFT description, which we show is impossible for a certain class of cosmological TCC violating backgrounds. Therefore, our argument goes beyond EFT and rests on quantum gravity features such as holography. 

Also, unlike the trans-Planckian problem, our reasoning does not apply to black holes. An essential difference between black hole backgrounds and the cosmological TCC-violating solutions considered in Sec. \ref{Sec:summary} is the choice of the observer.  If we consider the modes corresponding to the free-falling observer crossing the horizon, the modes are non-singular and the UV modes are in the vacuum. The trans-Planckian problem in black hole physics arises when one considers the infinite boost that relates measurements of the free-falling observer to those of an observer who hovers just above the horizon.  In fact, the same issue could be raised for Unruh radiation in Rindler spacetime which is a coordinate reparameterization of Minkowski spacetime. However, the effective field theory is perfectly trustable in the Minkowski background. 

In our argument, the check for EFT validity is whether the physical observables become trans-Planckian for  free-falling observers. If so, the EFT is invalid. In the case of black hole, this is not believed to happen. However, in our cosmological setting, it is indeed the free-falling observer that is stationary with respect to the comoving frame who measures trans-Planckian energy densities.  EFT is invalid because there is no frame in which the trans-Planckian issue is resolved.

\section{Implications for inflationary cosmology in the string theory landscape}\label{Sec:5}

In this section we discuss the implications of our findings for the interior of the string theory landscape where there is less perturbative control. Let us review some well-known observations about the string theory landscape that are relevant: 
\begin{enumerate}
    \item A generic string theory construction comes with multiple light scalar fields. Therefore, the scalar field potential often depends on several light scalar fields rather than one.
    \item The potential decreases or grows exponentially in the asymptotic directions of the scalar field space.
    \item The directions in which the scalar field potential decreases are steep enough to make any flat FRW solution decelerating in the future \cite{Rudelius:2021azq}. 
\end{enumerate}

Based on the above observations, one can see that a generic flat FRW solution which rolls down the potential has a decelerating future and a singular past in which the EFT breaks down due to large values of the scalar potential \cite{Bedroya:2022tbh}. Since the futures of these solutions are decelerating, they satisfy the assumptions we listed in Sec. \ref{Sec:summary} which makes the string theory landscape an excellent setup to apply our findings. We conclude the following:
\vspace{10pt}

\textit{\textbf{Implications for the string theory landscape: } Any classical expanding scalar field cosmology in the string theory landscape in which the scalar field rolls to the asymptotic of the field space must satisfy TCC. Note that we show TCC must be satisfied even if the scalar fields travel through the deep interior of the moduli space.}
\vspace{10pt}

To show the above conclusion, we showed that if TCC is violated, the EFT breaks down for the vacuum with zero particle excitations in the asymptotics of spacetime. However, this invalidates the EFT altogether given that in perturbative string theory, the EFT by definition is supposed to capture the physics of low-energy perturbations around that specific vacuum state.

One type of scalar potential that would violate TCC is a long stretch of a sufficiently long plateau that would allow $\ln(M_{\rm pl}/H)$ e-folds of quasi-de Sitter expansion. This is precisely the type of potential that is conventionally used in inflationary models. Our findings rule out the existence of such a plateau in the landscape as long as there is \text{any} classical solution that connects the quasi-de Sitter phase sourced by this plateau to the asymptotic region of the field space. In inflationary models, it is typically imagined that after inflation, the scalar field stabilizes at a local minimum of the potential. We emphasize that even if the scalar field does not roll to the asymptotic region of field space for one particular solution, the model can still be ruled out as long as \textit{any} such solution exists.

Our argument falls short of completely ruling out the inflationary scenario in string theory. This is due to the possibility that \textit{every} TCC-violating accelerating expanding solutions drain into a strictly local non-negative minimum or a negative minimum of the scalar potential, although different solutions can drain into different minima. These possibilities share the property that, for one reason or another, there is no classical solution which connects the TCC-violating phase to the asymptotic vacuum.\footnote{If the solution ends in a global minimum with zero potential energy, our arguments still apply due to eternal deceleration. If the solution goes to a negative minimum, it will ultimately contract to a singularity. If the ending point is a positive local minimum, it will eventually decay via quantum tunneling. A global positive minimum is known to be ruled out in string theory \cite{Obied:2018sgi}.}

Interestingly, for positive potentials it has proven difficult to construct either inflation (i.e. very flat potentials) or metastable de Sitter (i.e. positive local minimum) in string theory and no full construction for either one is currently known. It is understood that realizing either feature would require a finely-tuned cancellation between different contributions to the derivatives of the scalar potential. 

One of our main findings is that in theories with positive scalar potentials, realizing one of these fine-tunings (i.e. inflation) in the string theory landscape, requires realizing the other one as well (i.e. metastable de Sitter). Achieving both is already challenging in a single-field potential. In the typical string theory landscape, the potential depends on many fields, which complicates matters. This is because any flat stretch of potential can allow different accelerating solutions depending on the initial conditions for the direction of the scalar field. Our argument then implies that \textit{any} TCC-violating accelerating solution \textit{anywhere} in the landscape must end in a metastable de Sitter phase, whether relevant to the inflationary scenario or not. Thus realizing inflation in a typical string landscape requires an immeasurably complex and highly-nonlinear condition across the entire string landscape that may not be possible to satisfy.

\vspace{20pt}

\section*{Acknowledgments}

We would like to thank Austin Joyce, Hayden
Lee, Miguel Montero, LianTao Wang, and Cumrun Vafa for helpful discussions and valuable comments. AB is supported in part by the Simons Foundation grant number 654561 and by the Princeton Gravity Initiative at Princeton University. QL is supported in part by the NSF grant PHY-2210498 and PHY-2207584 and by the Simons Foundation. PJS is supported in part by the  Department of Energy grant number
DEFG02-91ER40671 and by the Simons Foundation grant number 654561.  PJS thanks
Cumrun Vafa and the High Energy Physics group in the Department of Physics at Harvard University for graciously hosting him during his sabbatical leave.

\appendix
\section{FRW mode expansions}\label{A1}

Suppose we have a theory with a scalar potential $V(\phi)$ which can be a function of multiple scalar fields with non-trivial scalar geometry. Consider a classical spatially-flat FRW solution of this theory that is solely driven by the scalar field potential and kinetic terms. Despite the presence of multiple scalar fields, we can parametrize the classical trajectory using a canonically normalized scalar field $\phi$. Therefore, this classical trajectory extremizes the following action.
\begin{align}\label{SFA}
   \int d^dx\sqrt{-g} [\frac{\mathcal{R}}{2\kappa}-\frac{1}{2}\partial_\mu \phi\partial^\mu \phi-V(\phi)]~,
\end{align}
where $\kappa=\sqrt{8\pi G}=\Mpl^{-(d-2)/2}$ is the reduced Planck factor. Note that we have excluded all the other matter fields As they would be irrelevant to our analysis. However, we will comment on the implicit assumption of weak coupling between $\phi$ and other fields in Sec. \ref{Sec:5}. Suppose the metric for the spatially-flat FRW solution is
\begin{align}
   ds^2= a(\tau)^2(-d\tau^2+\sum_{1\leq i \leq d-1}dx^idx^i),
\end{align}
where $\tau$ is the conformal time. We decompose the scalar field into the classical background $\phi_0(\tau)$ and fluctuation $\varphi$. 
\begin{align}
    \phi(\tau,{\bf{x}})=\phi_0(\tau)+\varphi(\tau,{\bf{x}}).
\end{align}
Substituting the above decomposition into the Lagrangian Eq. \eqref{SFA} leads to
\begin{align}
    \mathcal{L}_\text{matter}&=\left( \frac{a(\tau)^{d-2}}{2}\phi_0'^2-a(\tau)^dV(\phi_0)\right)\nonumber\\
    &+\left( a(\tau)^{d-2}\phi_0'\varphi'+\frac{a(\tau)^{d-2}}{2}\varphi'^2\right)\nonumber\\
    &-\left(\frac{a(\tau)^{d-2}}{2}(\vec\nabla\varphi)^2+a(\tau)^d   \partial_{\phi} V(\phi_0)  \varphi+a(\tau)^d\frac{\partial_\phi^2V(\phi_0)}{2}\varphi^2+\hdots\right),
\end{align}
where prime denotes derivative with respect to $\tau$. Note that since we will be promarily interested in the mode expansion for the scalar field and not graviton, we have excluded the gravitational term $\mathcal{R}$.

To define a vacuum we need to find the correct mode expansion.
\begin{align}\label{ME}
    \varphi(\tau,{\bf{x}})=\int \frac{d^{d-1}{\bf{k}}}{(2\pi)^{d-1}}\left[u(\tau,{\bf{k}})b({\bf{k}})+u^*(\tau,-{\bf{k}})b^\dagger(-{\bf{k}})\right]e^{i{\bf{k}}\cdot\bf{x}}\,,
\end{align}
where $u(\tau,{\bf{k}})e^{i{\bf{k}}\cdot\bf{x}}$ are solutions to the equation of motions and are normalized such that the creation annihilation operators $b^\dagger$ and $b$ satisfy the commutation relations
\begin{align}\label{BCR}
    [b({\bf{k}}_1),b^\dagger({\bf{k}}_2)]=(2\pi)^{d-1}\delta^{d-1}({\bf{k}}_1-{\bf{k}}_2)\,.
\end{align}
To see how the above condition imposes a normalization condition on $u(\tau,\bf{k})$, we must use the canonical commutation relations 
\begin{align}\label{CCR}
    [\varphi(\tau,{\bf{x}}),\pi(\tau,{\bf{y}})]=i\delta^{d-1}({\bf{x}}-{\bf{y}})\,,
\end{align}
where $\pi(\tau,{\bf{x}})$ is the canonical momentum conjugate given by 
\begin{align}
    \pi(\tau,{\bf{x}})=\frac{\delta \mathcal{L}}{\delta\varphi'}=a(\tau)^{d-2}(\phi_0'+\varphi')\,.
\end{align}
Generally, the commutation relations in non-inertial coordinate systems curvature dependent factors on the left side of \ref{CCR}. However, in the case of conformally flat spacetime such as ours, such factors are absent. To see why, suppose we were working in a coordinate system $t_\text{normal},x_\text{normal}^i$ which was instantaneously inertial at the point $(\tau,\bf{x})$. We can find such coordinates by ensuring the following differential equations are satisfied at the point of interest.
\begin{align}
    dt_\text{normal}&=a(\tau)d\tau\nonumber\\
    dx_\text{normal}&=a(\tau)dx^i\,.
\end{align}
Therefore, the canonical momentum conjugate $\pi$ and the delta functions are related as follows across the two coordinate systems. 
\begin{align}
    &\pi_{\text{normal}}=\partial_{t_{\text{normal}}}\phi_0+\partial_{t_\text{normal}}\varphi=a(\tau)^{-(d-1)}\pi\nonumber\\
    &\delta^{d-1}({\bf{x}-\bf{y}})_{\text{normal}}=\delta^{d-1}(a(\tau)({\bf{x}-\bf{y}}))=a(\tau)^{-(d-1)}\delta({\bf{x}-\bf{y}})\,.
\end{align}
Therefore, we find 
\begin{align}
    [\varphi(\tau,{\bf{x}}),\pi(\tau,{\bf{y}})]&= a(\tau)^{d-1}[\varphi(\tau,{\bf{x}}),\pi_{\text{normal}}(\tau,{\bf{y}})]\nonumber\\
    &=ia(\tau)^{d-1}\delta^{d-1}({\bf{x}-\bf{y}})_{\text{normal}}\nonumber\\
    &=i\delta({\bf{x}-\bf{y}})\,.
\end{align}
Note that we are working in natural units throughout this article. Using the mode expansion Eq. \eqref{ME} for $\varphi(\tau,{\bf{x}})$, we find
\begin{align}
     \pi(\tau,{\bf{x}})=a(\tau)^{d-2}\phi_0'+a(\tau)^{d-2}\int \frac{d^{d-1}{\bf{k}}}{(2\pi)^{d-1}}\left[u'(\tau,{\bf{k}})b({\bf{k}})+{u^*}'(\tau,-{\bf{k}})b^\dagger(-{\bf{k}})\right]e^{i{\bf{k}}\cdot\bf{x}}\,.
\end{align}
Note that the first term is proportional to the identity operator and therefore commutes with any operator. Therefore, we find
\begin{align}
     [\varphi(\tau,{\bf{x}}),\pi(\tau,{\bf{y}})]=&a(\tau)^{d-2}\int \frac{d^{d-1}{\bf{k}}}{(2\pi)^{d-1}}\int \frac{d^{d-1}{\bf{k}}'}{(2\pi)^{d-1}}e^{i({\bf{k}}\cdot\bf{x}+{\bf{k}}'\cdot\bf{y})}\nonumber\\
     &\left[u(\tau,{\bf{k}})b({\bf{k}})+u^*(\tau,-{\bf{k}})b^\dagger(-{\bf{k}}),u'(\tau,{\bf{k}}')b({\bf{k}}')+{u^*}'(\tau,-{\bf{k}}')b^\dagger(-{\bf{k}}')\right]\,.
\end{align}
Using the commutation relations Eq. \eqref{BCR} for $b({\bf{k}})$ and $b({\bf{k}})^\dagger$, we find
\begin{align}
     [\varphi(\tau,{\bf{x}}),\pi(\tau,{\bf{y}})]=a(\tau)^{d-2}\int \frac{d^{d-1}{\bf{k}}}{(2\pi)^{d-1}}e^{i{\bf{k}}\cdot({\bf{x}}-{\bf{y}})}\left[u(\tau,{\bf{k}}){u^*}'(\tau,{\bf{k}})-u'(\tau,-{\bf{k}})u^*(\tau,-{\bf{k}}))\right]\,.
\end{align}
To reproduce the canonical commutation relation Eq. \eqref{CCR}, we need
\begin{align}
    u(\tau,{\bf{k}}){u^*}'(\tau,{\bf{k}})-u'(\tau,-{\bf{k}})u^*(\tau,-{\bf{k}}))=ia(\tau)^{-(d-2)}\,.
\end{align}
The equation of motion $(\frac{\delta\mathcal{L}}{\delta\varphi}=0)$ that each mode satisfies is
\begin{align}
   [a(\tau)^d \partial_{\phi} V(\phi_0)+\partial_\tau(a(\tau)^{d-2}\phi_0')]+&[a(\tau)^d(\partial_{\phi}V(\phi_0+\varphi)-\partial_{\phi} V(\phi_0))\nonumber\\&+\partial_\tau(a(\tau)^{d-2}\varphi')-a(\tau)^{d-2}\nabla^2\varphi]=0\,.
\end{align}
Note that the first bracket vanishes because $\phi_0$ is a classical background that solves the equation of motion. So, we find
\begin{align}
    a(\tau)^2(\partial_{\phi}V(\phi_0+\varphi)-\partial_{\phi}V(\phi_0))+(d-2)\frac{a'}{a}\varphi'+\varphi''-\nabla^2\varphi=0\,.
\end{align}
By considering the leading term in the expansion of the potential we find
\begin{align}
    \partial_{\phi}V(\phi_0+\varphi)-\partial_\phi V(\phi_0)\simeq \varphi \partial_\phi^2 V(\phi_0)\,.
\end{align}
Substituting $\varphi=u_k(\tau)\cdot\exp(i{\bf{k}}\cdot{\bf{x}})$ in the approximate equation of motion leads to
\begin{align}
    (d-2)a(\tau)Hu_k'(\tau)+u_k''(\tau)+k^2u_k(\tau)+  a(\tau)^2\partial_\phi^2 V(\phi_0(\tau))u_k(\tau)\simeq 0\,,
\end{align}
where $H=a'/a^2$ is the Hubble parameter


The above equation has solutions that have positive and a negative frequencies at $\mathcal{I}^+$. To define the future vacuum, we take $u$ to be the positive frequency solution. 

\bibliographystyle{utphys}
    \bibliography{References}
\end{document}